\documentclass[11pt,a4paper]{article}
\pdfoutput=1
\usepackage{jheppub}
\usepackage[T1]{fontenc}
\usepackage[utf8]{inputenc}
\usepackage[UKenglish]{babel}
\usepackage{lmodern, feynmp, pdflscape}

\makeatletter\g@addto@macro\bfseries{\boldmath}\makeatother
\usepackage{ifpdf,feynmp}
\ifpdf\fi
\AtBeginDocument{
	\begin{fmffile}{graphs/graph}
	\fmfcmd{%
		prologues:=3;
		thin := .5pt;
		arrow_ang := 20;
		arrow_len := 3.5thick;
	}
}

\def\figureautorefname~#1\null{Fig.\,#1\null}

\def\equationautorefname~#1\null{Eq.\,(#1)\null}
\def\pageautorefname\nobreakspace{p.}

\makeatletter%
\renewcommand{\p@subsection}{\thesection.}
\newcommand*{\lcite }[1]{\hyper@@link[cite]{}{cite.#1}{\cite{#1}}}
\newcommand*{\lcites}[2]{\hyper@@link[cite]{}{cite.#1}{\cite{#2}}}
\newcommand*{\rcite }[1]{\hyper@@link[cite]{}{cite.#1}{Ref.\,\cite{#1}}}
\newcommand*{\rcites}[2]{\hyper@@link[cite]{}{cite.#1}{Refs.\,\cite{#2}}}
\makeatother%

\graphicspath{{figs/}}

\allowdisplaybreaks[4]
\DeclareMathOperator{\sign}{sign}
\let\Re\oldRe
\DeclareMathOperator{\Re}{Re}
\let\Im\oldIm
\DeclareMathOperator{\Im}{Im}
\newcommand*{\too}{\longrightarrow}

\newcommand{\re}[2]{\Re\!\big\{#1^*#2\big\}}
\newcommand{\im}[2]{\Im\!\big\{#1^*#2\big\}}
\renewcommand{\mod}[1]{\left|#1\right|^2}
\renewcommand{\a}[1]{%
	\if#11%
		A_+
	\else%
		A_-
	\fi}
\renewcommand{\b}[1]{%
	\if#11%
		B_+
	\else%
		B_-
	\fi}
\renewcommand{\c}[1]{%
	\if#11%
		C_+
	\else%
		C_-
	\fi}

\renewcommand*{\vec}[1]{\boldsymbol{#1}}
\newcommand*{\jpsi}{J\!/\!\psi}
\renewcommand*{\P}{\text{P}}
\newcommand*{\T}{\ensuremath{\hat{\text{T}}}}
\newcommand*{\CP}{\ensuremath{\text{CP}}}

\title{\CP\ violation in multibody decays of beauty baryons}

\newcommand{\desy}{DESY, Notkestrasse 85, D-22607 Hamburg, Germany}
\author{Gauthier Durieux}
\affiliation{\desy}

\abstract{
	Beauty baryons are being observed in large numbers in the LHCb detector.
	The rich kinematic distributions of their multibody decays are therefore
	becoming accessible and provide us with new opportunities to search for
	\CP\ violation. We analyse the angular distributions of some three- and
	four-body decays of spin-$1/2$ baryons using the Jacob--Wick helicity
	formalism. The asymmetries that provide access to small differences of
	\CP-odd phases between decay amplitudes of identical \CP-even phases are
	notably discussed. The understanding gained on processes featuring
	specific resonant intermediate states allows us to establish which
	asymmetries are relevant for what purpose. It is for instance shown that
	some \CP-odd angular asymmetries measured by the LHCb collaboration in
	the $\Lambda_b \to \Lambda\,\varphi \to p\,\pi\, K^+ K^-$ decay are
	expected to vanish identically.
}
\preprint{DESY 16-153}
\arxivnumber{}
\notoc
\makeatletter\renewcommand{\@fpheader}{}\makeatother

\begin{document}
\maketitle
\vfill
\clearpage

\section{Introduction}
Despite of production rates somewhat smaller than that of mesons, beauty baryons
are now being observed in significant numbers in the LHCb detector. They have
therefore started to offer complementary means to test the standard model. The
search for new sources of \CP\ violation is an especially relevant direction in
which they could provide new opportunities. Incidentally, a first hint of \CP\
violation could just have been observed in the $\Lambda_b\to p\,\pi^-\pi^+\pi^-$
channel~\lcites{Vieites:2016ichep}{Vieites:2016ichep, *Aaij:2016cla}. Using
angular momentum conservation through
the Jacob--Wick helicity formalism, we aim at determining what angular
asymmetries can be expected in specific beauty baryon decays as well as how they
relate to the underlying dynamics and its discrete symmetry properties. By
discussing the case of spin-$1/2$ baryon decays, this paper extends
\rcite{Durieux:2015zwa} that focused on the case of spin-$0$ particles. Most of
our results also apply to the decay of any spin-$1/2$ state.

A violation of \CP, sourced in the standard model or beyond, manifests itself
through relative \CP-odd phases---also called \emph{weak phases}---between decay
amplitudes. They can be accessed through interferences in which \CP-even---or
\emph{strong}---phases originating from the absorptive parts of amplitudes can
also appear. The most common interferences take the following form:
\begin{align*}
	\Re\{A^*_1A_2\}
	&=
	|A^*_1A_2|\; \Re\big\{ e^{i\Delta\delta_{12}+i\Delta\varphi_{12}} \big\}
	\\&=
	|A^*_1A_2|\, \big(
		 \cos\Delta\delta_{12}\:	\cos\Delta\varphi_{12}
		-\sin\Delta\delta_{12}\:	\sin\Delta\varphi_{12}
		\big),
\end{align*}
where $\Delta\varphi$ and $\Delta\delta$ respectively denote \CP-odd and
\CP-even phase differences. The second \CP-odd term can be extracted by
combining \CP-conjugate processes, through rate asymmetries notably. It provides
sensitivity to small differences of weak phases, a sensitivity which is however
conditioned on the presence of relative strong phases. Some other interferences
take the
\begin{equation*}
	\Im\{A^*_1A_2\}=
	|A^*_1A_2|\, \big(
		 \sin\Delta\delta_{12}\:	\cos\Delta\varphi_{12}
		+\cos\Delta\delta_{12}\:	\sin\Delta\varphi_{12}
		\big)
\end{equation*}
form. The second \CP-odd term extracted by combining \CP-conjugate processes
is again sensitive to small weak phase differences,
but does not vanish in the absence of relative strong phases. Studying this
second type of interferences is therefore particularly relevant in cases where
small strong phases are expected. Measuring both types of interferences can also 
lead to a better understanding of strong phases which are difficult to compute
when they result from nonperturbative dynamics.

Beside rate asymmetries already mentioned, differential distributions can serve
to access various interference terms. Exploiting the distributions of decay
products instead of decay rates can also be advantageous when the production
cross sections of \CP-conjugate particles differ---as they generally do in $pp$
collisions---and production rate asymmetries are not precisely known. It is
useful to define motion reversal \T\ (often called \emph{naive time reversal}),
a transformation that reverts momentum and spin three-vectors. Indeed, the
motion reversal properties of differential distributions determine which type of
amplitude interferences they give access to: \T-even observables provide access
to the $\Re\{A^*_iA_j\}$ interferences, \T-odd observables to the
$\Im\{A^*_iA_j\}$ ones. Let us focus somewhat on the \T-odd observables which
thus yield sensitivity to small differences of \CP-odd phases between amplitudes
having small or vanishing relative \CP-even phases. In a Lorentz-invariant form,
\T-odd variables only appear proportional to a completely antisymmetric
$\epsilon_{\mu\nu\rho\sigma}$ contraction of four independent four-vectors. In
processes involving only spinless external states, they can thus only be
constructed when at least five external particles are involved, like in
four-body decays. In processes involving spinning particles, \T-odd variables
can in principle also be constructed through the antisymmetric contraction of
both momentum and spin four-vectors. They constitute qualitatively different
observables.
Unlike momenta, the spin vectors of stable particles are however practically
unmeasurable in the context we are interested in. So we will refrain from
considering as observables the $\epsilon_{\mu\nu\rho\sigma}$ contractions in
which they appear (that give rise to triple products like $\vec s\cdot (\vec p_i
\times \vec p_j)$ in a specific frame). Only angular distributions that derive
from measured final-state momenta will be awarded that status. Final-state spins
will be altogether disregarded and summed over. The polarisation of the decaying
particle can however be considered as resulting from the production process since it
is determined by production amplitudes. In the decay of spinning particles, the
angular distributions of decay products can then be viewed as providing access to
combinations of production and decay amplitudes. This is to be contrasted
with the decays of spinless particles where they provide direct
access to decay amplitudes.

From this more practical point of view, here is how spinning particles offer new
opportunities to search for small differences of \CP-odd phases between decay
amplitudes that have identical---potentially vanishing---\CP-even phases. As a
matter of fact, \T-even angular distributions still provide access to small
\CP-odd phase differences only in the presence of relative \CP-even phases. The
latter can however appear in the production amplitudes, as angular distributions
now give access to an entwined combination of production and decay amplitudes.
Such strong phases in production amplitudes would manifest themselves as a
nonvanishing \T-odd polarisation component, which we will denote $P_z$. As a
result, certain imaginary parts of decay amplitude interferences
become accessible through \T-even angular distributions, in terms proportional
to this \T-odd polarisation component of the decaying particle.
In particular, there are not enough independent external-particle four-momenta
in three-body decays to form an antisymmetric $\epsilon_{\mu\nu\rho\sigma}$
contraction. One must necessarily rely on at least one spin four-vector to form
a \T-odd variable. As will be illustrated below with final-state spins summed
over, the imaginary parts of decay amplitude interferences then only appear in
terms proportional to the decaying particle polarisation.
A positive signal of \CP\ violation in one of the corresponding asymmetries could
thus be sourced either in decay amplitudes or in production ones, leading, in
the latter case, to a mismatch between the polarisation of the initial particle
and minus the polarisation of its antiparticle. Such an effect is not expected
to be sizeable when the strong interaction which conserves \CP\ dominates the
production process. Without assuming it is altogether absent, one would have to
rely on a comparison between the expected and measured patterns of asymmetries
to discriminate between these two possibilities. The patterns expected for
decays through specific resonant intermediate states are presented below.

On the other hand, all the \T-odd angular distributions can no longer serve to
isolate small differences of \CP-odd phases between decay
amplitudes of identical \CP-even phases. The \T-odd angular distributions that
appear proportional to $P_z$ no longer give access to imaginary parts of decay
amplitude interferences. Sensitivity to hypothetical \CP-odd phase differences
through these terms then actually relies on the presence of nonvanishing
\CP-even phase differences between the corresponding decay amplitudes.
A systematic and \emph{blind} construction of \T-odd--\CP-odd asymmetries, as
performed in \rcite{Durieux:2015zwa} for the decay of spinless particles, is
nevertheless possible. We stress this procedure can still be utilized experimentally to
cover the unexpected or in situations where complicated patterns of
interferences are not described precisely enough. One would however need to rely
on specific results such as the ones presented here for selected resonance
structures to establish whether a given \T-odd angular asymmetry yields
sensitivity to \CP-odd phase differences between production or decay amplitudes.

Aside from \CP\ violation, one can in principle measure all prescribed
independent contributions to the angular distributions and thereby gain
further understanding about the process under scrutiny. Our tables establish the
necessary link between kinematic distributions and the dynamics encoded in
amplitudes. The precision achieved will obviously depend on the collected
statistics, but note the determination of each asymmetry or moment exploits the
statistical power of the full data sample. This is to be contrasted with a fit
in which additional free parameters worsen the precision to which all of them
can be determined (see e.g. \rcite{Beaujean:2015xea}).

\begin{figure}[t]\center
\begin{tabular}{l@{\hspace{2cm}}l}
	\fmfframe(2,2)(5,2){
	\begin{fmfgraph*}(30,17)
	\fmfleft{l}
	\fmfright{r1,r2,r3,r4}
	\fmf{vanilla,tens=4}{l,v}	\fmflabel{$0^{1/2}$}{l}
	\fmf{vanilla,tens=2,lab=$b^1$, lab.side=right}{v,w1}
	\fmf{vanilla,tens=2,lab=$a^{1/2,,3/2}$, lab.side=left}{v,w2}
	\fmf{vanilla}{r1,w1,r2}	\fmflabel{$4^{1/2}$}{r1}	\fmflabel{$3^{1/2}$}{r2}
	\fmf{vanilla}{r3,w2,r4}	\fmflabel{$2^{0}$}{r3}		\fmflabel{$1^{1/2}$}{r4}
	\end{fmfgraph*}
	}
	&
	\fmfframe(2,2)(2,2){
	\begin{fmfgraph*}(30,17)
	\fmfleft{l}
	\fmfright{r1,r2,r3,r4}
	\fmf{vanilla,tens=4}{l,v}	\fmflabel{$0^{1/2}$}{l}
	\fmf{vanilla,tens=2,lab=$b^1$, lab.side=right}{v,w1}
	\fmf{vanilla,tens=2,lab=$a^{1/2,,3/2}$, lab.side=left}{v,w2}
	\fmf{vanilla}{r1,w1,r2}	\fmflabel{$4^{0}$}{r1}	\fmflabel{$3^{0}$}{r2}
	\fmf{vanilla}{r3,w2,r4}	\fmflabel{$2^{0}$}{r3}		\fmflabel{$1^{1/2}$}{r4}
	\end{fmfgraph*}
	}
	\\[10mm]
	\fmfframe(2,2)(5,2){
	\begin{fmfgraph*}(30,17)
	\fmfleft{l}
	\fmfright{r1,r2,r3}
	\fmf{vanilla,tens=3}{l,v}	\fmflabel{$0^{1/2}$}{l}
	\fmf{vanilla,tens=2,lab=$a^{1/2,,3/2}$, lab.side=left}{v,w}
	\fmf{vanilla}{r2,w,r3}		\fmflabel{$2^{0}$}{r2}		\fmflabel{$1^{1/2}$}{r3}
	\fmf{vanilla}{v,r1}		\fmflabel{$b^{1}$}{r1}
	\end{fmfgraph*}
	}
	&
	\fmfframe(2,2)(5,2){
	\begin{fmfgraph*}(30,17)
	\fmfleft{l}
	\fmfright{r1,r2,r3}
	\fmf{vanilla,tens=3}{l,v}	\fmflabel{$0^{1/2}$}{l}
	\fmf{vanilla,tens=2,lab=$a^{1/2,,3/2}$, lab.side=left}{v,w}
	\fmf{vanilla}{r2,w,r3}		\fmflabel{$2^{0}$}{r2}		\fmflabel{$1^{1/2}$}{r3}
	\fmf{vanilla}{v,r1}		\fmflabel{$b^{0}$}{r1}
	\end{fmfgraph*}
	}
\end{tabular}
\caption{The eight three- and four-body decays considered in this paper. The
         superscripts to particles' labels specify their spins.}
\label{fig:proc}
\end{figure}
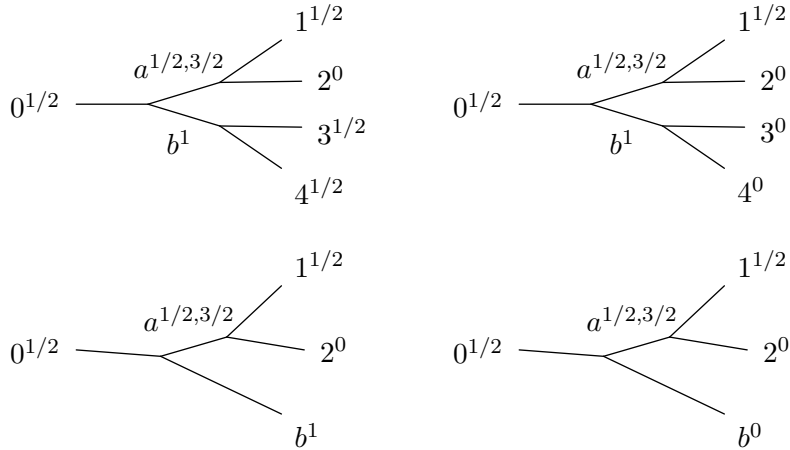

\begin{figure}[t]\center
\includegraphics[scale=1.25]{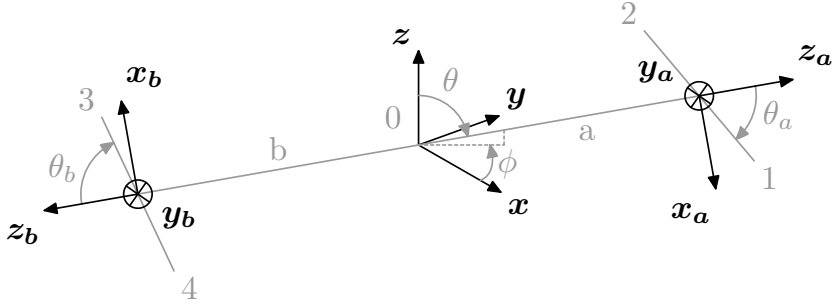}
\caption{Reference frames defined according to the Jackson
         convention~\protect\lcite{Jackson:1965gda} where axes in the two
         daughter restframes are (anti)aligned. The azimuthal angles
         $\phi_{a,b}$ that are not apparent are defined in the usual way:
         measured from the $\vec{x_{a,b}}$ axes such that the $\vec{y_{a,b}}$
         axes have $\phi_{a,b}=+\pi/2$. Note the $a,b$ particles' momenta are
         pictured in the $0$ particle restframe, while the $1,2$ and $3,4$ ones
         in the $a$ and $b$ restframes, respectively.}
\label{fig:frames}
\end{figure}

\section{Angular distributions}
The following four- and three-body
decays (depicted in \autoref{fig:proc}) will be considered:
\begin{equation*}
\renewcommand{\arraystretch}{1.2}
	\begin{array}{*{3}{l}}
	0^{1/2}	&\too a^{1/2,3/2}\;b^{1} 
			&\too 1^{1/2}\,2^{0} \; 3^{1/2}\,4^{1/2}	,\\
		&	&\too 1^{1/2}\,2^{0} \; 3^{0}\,4^{0}		,
	\end{array}
	\hspace{1.75cm}
	\begin{array}{*{3}{l}}
	0^{1/2}	&\too a^{1/2,3/2}\;b^{1}
			&\too 1^{1/2}\,2^{0} \; b^{1}		,\\
		&\too a^{1/2,3/2}\;b^{0}
			&\too 1^{1/2}\,2^{0} \; b^{0}		,
	\end{array}
\end{equation*}
where the superscripts of particle labels specify their spins. Examples of such
processes include the $\Lambda_b\to \Lambda \, \jpsi \to p\,\pi\, \mu^+\mu^-$,
$\Lambda_b\to \Lambda\, \varphi\to p\,\pi\, K^+K^-$, $\Lambda_b\to N^*K_s\to
p\pi K_s$ decays which were studied by the LHCb collaboration in the recent
\rcites{Aaij:2013oxa}{Aaij:2013oxa, Aaij:2016zhm, Aaij:2014lpa}, or the
$\Lambda_b\to \Lambda(X)\gamma \to p \,K\!/\!\pi\, \gamma$ processes discussed
in \rcites{Gremm:1995nx}{Gremm:1995nx, Mannel:1997xy, Hiller:2001zj,
Legger:2006cq, Hiller:2007ur}. Standard-model predictions relying on
factorization for some charmless multibody hadronic $b$-baryon decays have been
discussed in \rcite{Bensalem:2002pz}. \T-odd--\CP-odd asymmetries are estimated
to reach approximately the twenty-percent level in $\Lambda_b\to N(1440) K^-$ or
$\Xi_b\to \Sigma^+ K^-$, the percent level in the $\Lambda_b\to N(1440) K^{*-}$
or $\Xi_b\to \Sigma^+ K^{*-}$, and the sub-percent level in $\Lambda_b\to
\Lambda \eta^{(\prime)}$ or $\Lambda_b\to \Lambda \varphi$ which are dominated by
one single penguin amplitude. In \rcite{Bensalem:2002ys}, it was argued that
new-physics contributions parametrized by effective operators (arising, e.g.,
from $Z'$ or R-parity-violating supersymmetry) could be significantly larger.
Additionally, the rate asymmetry in $\Lambda_b\to \Lambda \gamma$ was
also estimated to reach at most the percent level in \rcite{Hiller:2001zj}.

The helicity formalism of Jakob and Wick~\lcite{Jacob:1959at} will be employed,
following the so-called \emph{Jackson convention}~\lcite{Jackson:1965gda} for
the definition of the various reference frames (see \autoref{fig:frames}). The
spins of final-state particles will be summed over. On the other hand, a
nonvanishing polarisation of the initial spin-$1/2$ baryon will be considered.
Although experimental datasets never isolate perfectly one single resonant
intermediate state, the interferences between them lie beyond the scope of this
work. Neither will topologies like
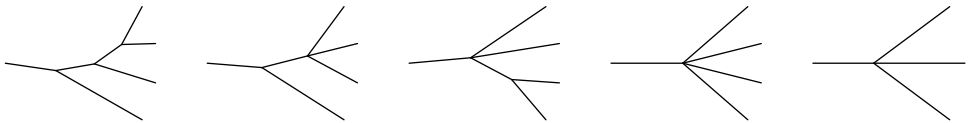
\begin{center}
\newcommand{\myfmfframe}{\fmfframe(2,1)(2,-1)}
	\myfmfframe{
	\begin{fmfgraph*}(20,15)
	\fmfleft{l}
	\fmfright{r1,r2,r3,r4}
	\fmf{vanilla,tens=4}{l,v1}
	\fmf{vanilla,tens=3}{v1,v2}
	\fmf{vanilla,tens=2}{v2,v3}
	\fmf{vanilla}{r3,v3,r4}
	\fmf{vanilla}{v2,r2}
	\fmf{vanilla}{v1,r1}
	\end{fmfgraph*}
	}
	\myfmfframe{
	\begin{fmfgraph*}(20,15)
	\fmfleft{l}
	\fmfright{r1,r2,r3,r4}
	\fmf{vanilla,tens=4}{l,v1}
	\fmf{vanilla,tens=3}{v1,v2}
	\fmf{vanilla}{r3,v2,r4}
	\fmf{vanilla}{v2,r2}
	\fmf{vanilla}{v1,r1}
	\end{fmfgraph*}
	}
	\myfmfframe{
	\begin{fmfgraph*}(20,15)
	\fmfleft{l}
	\fmfright{r1,r2,r3,r4}
	\fmf{vanilla,tens=4}{l,v1}
	\fmf{vanilla,tens=2}{v1,v2}
	\fmf{vanilla}{r3,v1,r4}
	\fmf{vanilla}{r1,v2,r2}
	\end{fmfgraph*}
	}
	\myfmfframe{
	\begin{fmfgraph*}(20,15)
	\fmfleft{l}
	\fmfright{r1,r2,r3,r4}
	\fmf{vanilla,tens=4}{l,v1}
	\fmf{vanilla}{r3,v1,r4}
	\fmf{vanilla}{r1,v1,r2}
	\end{fmfgraph*}
	}
	\myfmfframe{
	\begin{fmfgraph*}(20,15)
	\fmfleft{l}
	\fmfright{r1,r2,r3}
	\fmf{vanilla,tens=4}{l,v1}
	\fmf{vanilla}{r3,v1}
	\fmf{vanilla}{r1,v1,r2}
	\end{fmfgraph*}
	}
\end{center}
be considered.

A first $(\vec x,\vec y,\vec z)$ system of axes is defined in the restframe of
the initial---mother---particle~$0$. When the production of the latter
preserves parity, its polarisation vector is orthogonal to the production plane
(see Sec.\,V of \rcite{Jacob:1959at}). To take advantage of this feature,
the $\vec z$ axis is taken parallel to the normal of the production plane. So is
a \emph{transversity frame} obtained.\footnote{Alternatively, a \emph{helicity
frame} could have been obtained with $\vec z$ aligned to particle~$0$'s momentum
in the laboratory frame (see \rcites{Conte:2008ema}{Conte:2008ema,
Leitner:2006nb, Leitner:2006sc} for some
results obtained in such a frame).} Particles $a$ and $b$ are respectively
produced at polar angles $\theta$ and $\pi-\theta$ from that $\vec{z}$ axis.
With the spin vector of particle $0$ pointing exactly in the $\vec z$ direction,
no dependence on the azimuthal angle $\phi$ of particle $a$ is
generated. The direction of the $\vec x$ axis is therefore chosen arbitrarily in
the plane perpendicular to the $\vec z$ axis.

Two other systems of axes are defined in the restframes of the $a$ and $b$
daughters, as so-called \emph{helicity frames}. The second one will only be
relevant for four-body decays. The $(\vec{x_a},\vec{y_a},\vec{z_a})$ system is
obtained by a $R(\phi,\theta,0)^T$ Euler rotation\footnote{A
$R(\phi,\theta,\chi)$ transformation is the succession of three elementary
rotations around the $\vec z$, $\vec y$, and $\vec z$ axes: $R_{\vec z}(\phi)
R_{\vec y}(\theta) R_{\vec z}(\chi)$. In the so-called \emph{Jacob--Wick
convention}, the $(\vec{x_a},\vec{y_a},\vec{z_a})$ frame is obtained by a
$R(\phi,\theta,-\phi)^T$ rotation of the $(\vec x,\vec y,\vec z)$ one.} of the
initial $(\vec x,\vec y,\vec z)$ one, followed by a suitable boost in the
$\vec{z_a}$ direction (parallel to particle~$a$'s momentum in the mother
restframe). A $R(\phi+\pi, \pi-\theta,0)^T$ rotation followed by a boost in the
$\vec{z_b}$ direction is required to obtain the
$(\vec{x_b},\vec{y_b},\vec{z_b})$ system. Its axes are parallel or antiparallel
to the $(\vec{x_a},\vec{y_a},\vec{z_a})$ ones.

The following assumptions will be made in the main text and relaxed in
\autoref{sec:appendix}:
\begin{itemize}
\item
	The production of particle $0$ preserves parity, so that its
	polarisation (if any) is aligned with the $\vec z$ axis. One therefore
	has $P_x=0=P_y$ in the density matrix for particle $0^{1/2}$:
	\begin{equation*}
	\newcommand*{\se}{\scriptstyle}
		\rho(m_0,m_0') =
			\frac{1}{2}\quad
			\bordermatrix{
				& \se+1/2	& \se-1/2	\cr
			\se+1/2	& 1+P_z		& P_x-iP_y \cr
			\se-1/2	& P_x+iP_y	& 1-P_z
			}
	\end{equation*}
	where $m_0^{(\prime)}$ is the component of particle $0$'s spin along the
	$\vec z$ axis.
\item
	When appearing as a final-state particle, in the three-body decays we
	consider, the $b^1$ vector is taken massless so that it has no
	$\lambda_b=0$ zero helicity state (and the $A_\pm$ amplitudes defined
	below are absent).
\item
	The $b^{1}\to3^{1/2}4^{1/2}$ decay preserves parity, so that its
	helicity amplitudes satisfy 
	\begin{equation*}
	|M_b(-\lambda_3, -\lambda_4)|^2 =
	|M_b(+\lambda_3, +\lambda_4)|^2.
	\end{equation*}
\item
	The $3^{1/2}$ and $4^{1/2}$ particles arising from the $b^1$ vector
	decay are massless and therefore have opposite helicities: $\lambda_3 =
	-\lambda_4$.
\end{itemize}
In each four-body process considered here, for the $b^{1}$ decay, there is
therefore one single independent combination of squared helicity amplitudes:
\begin{equation*}
	\text{either}\quad
		|M_b(1/2,-1/2)|^2+|M_b(-1/2,1/2)|^2
	,
	\quad\text{or}\quad
	|M_b(0,0)|^2.
\end{equation*}
They will be absorbed into the definition of the $M_0$ helicity amplitudes for
the parent $0\to a\, b$ decay. We will also absorb in the $M_0$'s the
$|M_a(+1/2,0)|^2+|M_a(-1/2,0)|^2$ combination of $a^{1/2,3/2}\to 1^{1/2}\,2^{0}$
amplitudes and define the
\begin{equation*}
\alpha_a \equiv \frac
	{|M_a(+1/2,0)|^2-|M_a(-1/2,0)|^2}
	{|M_a(+1/2,0)|^2+|M_a(-1/2,0)|^2}
\end{equation*}
asymmetry parameter which violates parity \P. It is therefore expected to vanish
if the corresponding decay proceeds through the strong interaction, like in the
$\Lambda(1520)\to pK$ example of $a^{3/2}\to 1^{1/2}2^0$ decay. The helicity
combinations allowed for the $a,b$ system are $(\lambda_a,\lambda_b)=(\pm
1/2,0)$, and $(\pm1/2,\pm1)$ for a spin-$1/2$ particle $a^{1/2}$, as well as
$(\pm3/2,\pm1)$ for a spin-$3/2$ particle $a^{3/2}$. We will denote the
corresponding amplitudes as
\begin{equation*}
	A_\pm \equiv M_0(\pm1/2,0)	,
	\qquad
	B_\pm \equiv M_0(\pm1/2,\pm1)	,
	\qquad
	C_\pm \equiv M_0(\pm3/2,\pm1)	.
\end{equation*}
As opposed to the rates of the four-body decays featuring $b^1$ as an
intermediate particle, the three-body ones in which it appears in the final
state will only contain interferences between amplitudes of identical
$\lambda_b$. Note also that a massless $b^1$ vector produced onshell can only
have $\lambda_b=\pm1$. The $A_\pm$ amplitudes therefore vanish in that case. As
already mentioned, this will be assumed in the main text for the three-body
$0^{1/2} \to a^{1/2,3/2}\;b^{1} \to 1^{1/2}\,2^{0} \; b^{1}$ decays. Finally,
beyond the narrow width approximation for particles $a$ and $b$, the
$A_\pm,B_\pm,C_\pm$ as well as $\alpha_a$ amplitudes have a non-trivial
dependence on the $(12)$ and $(34)$ invariant masses which we will respectively
denote $m_a$ and $m_b$.

The various contributions to the $\text{d}\Gamma/\text{d}\Omega$ angular
distributions of the final-state particles for the processes depicted in
\autoref{fig:proc} are given in \autoref{tab:proc1}--\ref{tab:proc_3body_4}.
The overall normalisation is chosen such that the
\begin{equation*}
	\int\text{d}\Omega \equiv\quad
	\int\limits_{-1}^{+1}\frac{\text{d}\cos\theta}{2}
	\int\limits_{-1}^{+1}\frac{\text{d}\cos\theta_a}{2}
	\int\limits_{-1}^{+1}\frac{\text{d}\cos\theta_b}{2}
	\int\limits_{-\pi}^{+\pi}\frac{\text{d}\phi}{2\pi}
	\int\limits_{-\pi}^{+\pi}\frac{\text{d}\phi_a}{2\pi}
	\int\limits_{-\pi}^{+\pi}\frac{\text{d}\phi_b}{2\pi}
\end{equation*}
angular integration simply yields the sum of the allowed amplitudes squared. In
case all the six of them are present, one would then get
\begin{equation*}
	\int\text{d}\Omega
	\frac{\text{d}\Gamma}{\text{d}\Omega} = 
	|\a1|^2 + |\a2|^2 + |\b1|^2 + |\b2|^2 + |\c1|^2 + |\c2|^2.
\end{equation*}
Here, again, the dependence on the $m_{a,b}$ invariant masses is kept implicit.

Some of the angular distributions we obtained have already been presented
elsewhere, sometimes partially only. \autoref{tab:proc1} agrees with the Table~1
of \rcite{Hrivnac:1994jx}. \autoref{tab:proc_3body_1} agrees with Eq.\,(7)
of \rcite{Legger:2006cq}. The terms in \autoref{tab:proc_3body_2} that are
not proportional to $\alpha_a$ only match Eq.\,(15) of
\rcite{Legger:2006cq}, provided $w_5$ and $w_6$ defined there are
respectively multiplied by factors of $\pm P_z$. The relative sign between
these $B_+C_+$ and $B_-C_-$ interferences can be understood given the
$d^j_{-\mu,-\lambda}(\theta)=(-1)^{\lambda-\mu} d^j_{\mu,\lambda}(\theta)$ symmetry
relations between Wigner matrices (see Eq.\,(A1) of \rcite{Jacob:1959at}).
Note also a relative complex conjugation of the amplitudes defined here and
there, as well as the use of the Jacob--Wick convention there which leads to
expressions identical to the ones we obtain with the Jackson convention for the
terms compared when $\phi_\Lambda$ is set to $0$ there. Both
\autoref{tab:proc_3body_3} and the $A_\pm$ dependence of \autoref{tab:proc2}
agree with Eq.\,(16) and (21) of \rcite{Kang:2010td}, obtained with
$P_z=+1$.

\begin{table}\center
\ensuremath{%
	\begin{array}{@{}  *{7}{c@{\quad}}  c@{}}
	\hline\\[-3mm]
	+ 3/2 		&\mod{\a1} + \mod{\a2}	& 	& 	& 	& 	&\sin^2\theta_b	& 	\\
+ 3/4 		&\mod{\b1} + \mod{\b2}	& 	& 	& 	& 	&1 + \cos^2\theta_b	& 	\\
+ 3/2 		&\mod{\a1} - \mod{\a2}	&\alpha_a	& 	& 	&\cos\theta_a	&\sin^2\theta_b	& 	\\
+ 3/4 		&\mod{\b1} - \mod{\b2}	&\alpha_a	& 	& 	&\cos\theta_a	&1 + \cos^2\theta_b	& 	\\
- 3\big/2\sqrt2 	&\re{\a1}{\b2} - \re{\a2}{\b1}	&\alpha_a	& 	& 	&\sin\theta_a	&\sin2\theta_b	&\cos(\phi_a + \phi_b)	\\
\\[-3mm]\hline\\[-3mm]
+ 3/2 		&\mod{\a1} - \mod{\a2}	& 	&P_z	&\cos\theta	& 	&\sin^2\theta_b	& 	\\
- 3/4 		&\mod{\b1} - \mod{\b2}	& 	&P_z	&\cos\theta	& 	&1 + \cos^2\theta_b	& 	\\
- 3\big/2\sqrt2 	&\re{\a1}{\b1} - \re{\a2}{\b2}	& 	&P_z	&\sin\theta	& 	&\sin2\theta_b	&\cos\phi_b	\\
+ 3/2 		&\mod{\a1} + \mod{\a2}	&\alpha_a	&P_z	&\cos\theta	&\cos\theta_a	&\sin^2\theta_b	& 	\\
- 3/4 		&\mod{\b1} + \mod{\b2}	&\alpha_a	&P_z	&\cos\theta	&\cos\theta_a	&1 + \cos^2\theta_b	& 	\\
- 3\big/2\sqrt2 	&\re{\a1}{\b2} + \re{\a2}{\b1}	&\alpha_a	&P_z	&\cos\theta	&\sin\theta_a	&\sin2\theta_b	&\cos(\phi_a + \phi_b)	\\
- 3\big/2\sqrt2 	&\re{\a1}{\b1} + \re{\a2}{\b2}	&\alpha_a	&P_z	&\sin\theta	&\cos\theta_a	&\sin2\theta_b	&\cos\phi_b	\\
- 3 		&\re{\a1}{\a2}	&\alpha_a	&P_z	&\sin\theta	&\sin\theta_a	&\sin^2\theta_b	&\cos\phi_a	\\
- 3/2 		&\re{\b1}{\b2}	&\alpha_a	&P_z	&\sin\theta	&\sin\theta_a	&\sin^2\theta_b	&\cos(\phi_a + 2 \phi_b)	\\
\\[-3mm]\hline\\[-3mm]
+ 3\big/2\sqrt2 	&\im{\a1}{\b1} + \im{\a2}{\b2}	& 	&P_z	&\sin\theta	& 	&\sin2\theta_b	&\sin\phi_b	\\
- 3\big/2\sqrt2 	&\im{\a1}{\b2} - \im{\a2}{\b1}	&\alpha_a	&P_z	&\cos\theta	&\sin\theta_a	&\sin2\theta_b	&\sin(\phi_a + \phi_b)	\\
+ 3\big/2\sqrt2 	&\im{\a1}{\b1} - \im{\a2}{\b2}	&\alpha_a	&P_z	&\sin\theta	&\cos\theta_a	&\sin2\theta_b	&\sin\phi_b	\\
- 3 		&\im{\a1}{\a2}	&\alpha_a	&P_z	&\sin\theta	&\sin\theta_a	&\sin^2\theta_b	&\sin\phi_a	\\
- 3/2 		&\im{\b1}{\b2}	&\alpha_a	&P_z	&\sin\theta	&\sin\theta_a	&\sin^2\theta_b	&\sin(\phi_a + 2 \phi_b)	\\
\\[-3mm]\hline\\[-3mm]
- 3\big/2\sqrt2 	&\im{\a1}{\b2} + \im{\a2}{\b1}	&\alpha_a	& 	& 	&\sin\theta_a	&\sin2\theta_b	&\sin(\phi_a + \phi_b)	\\

	\\[-3mm]\hline
	\end{array}%
}
\caption{Various contributions to the angular distribution of the $0^{1/2}\to
         a^{1/2}b^{1}\to 1^{1/2}\,2^{0}\:3^{1/2}\,4^{1/2}$ process, with
         conventions and assumptions specified in the text. Each line
         corresponds to a term of different angular dependence (most of them
         being independent). The separation in columns is only meant to ease the
         comparison between the various factors appearing in each term. The four
         blocks distinguish terms whose combinations of angular and polarisation
         dependence have different parity and motion reversal transformation
         properties. They are respectively \P-even--\T-even, \P-odd--\T-even,
         \P-even--\T-odd, and \P-odd--\T-odd.}
\label{tab:proc1}
\end{table}

\begin{table}\center
\vspace{-1.3cm}
\scalebox{.825}{%
\ensuremath{%
	\begin{array}{@{}  *{7}{c@{\;\;}}  c@{}}
	\hline\\[-3mm]
	+ 3/8 		&\mod{\b1} + \mod{\b2} + 3 \mod{\c1} + 3 \mod{\c2}	& 	& 	& 	& 	&1 + \cos^2\theta_b	& 	\\
- 3\sqrt{2/3}\big/4 	&\re{\a1}{\c1} + \re{\a2}{\c2}	& 	& 	& 	&\sin2\theta_a	&\sin2\theta_b	&\cos(\phi_a + \phi_b)	\\
+ 3/4 		&\mod{\a1} + \mod{\a2}	& 	& 	& 	&(1 + 3 \cos^2\theta_a)	&\sin^2\theta_b	& 	\\
- 3\sqrt3\big/4 	&\re{\b1}{\c2} + \re{\b2}{\c1}	& 	& 	& 	&\sin^2\theta_a	&\sin^2\theta_b	&\cos(2 \phi_a + 2 \phi_b)	\\
+ 9/8 		&\mod{\b1} + \mod{\b2} - \mod{\c1} - \mod{\c2}	& 	& 	& 	&\cos^2\theta_a	&1 + \cos^2\theta_b	& 	\\
- 3/8 		&5 \mod{\b1} - 5 \mod{\b2} - 3 \mod{\c1} + 3 \mod{\c2}	&\alpha_a	& 	& 	&\cos\theta_a	&1 + \cos^2\theta_b	& 	\\
- 3/4 		&\mod{\a1} - \mod{\a2}	&\alpha_a	& 	& 	&(5 - 9 \cos^2\theta_a) \cos\theta_a	&\sin^2\theta_b	& 	\\
+ 3\sqrt{2/3}\big/4 	&\re{\a1}{\c1} - \re{\a2}{\c2}	&\alpha_a	& 	& 	&(1 - 3 \cos^2\theta_a) \sin\theta_a	&\sin2\theta_b	&\cos(\phi_a + \phi_b)	\\
+ 3\big/4\sqrt2 	&\re{\a1}{\b2} - \re{\a2}{\b1}	&\alpha_a	& 	& 	&(1 - 9 \cos^2\theta_a) \sin\theta_a	&\sin2\theta_b	&\cos(\phi_a + \phi_b)	\\
+ 9\sqrt3\big/4 	&\re{\b1}{\c2} - \re{\b2}{\c1}	&\alpha_a	& 	& 	&\sin^2\theta_a \cos\theta_a	&\sin^2\theta_b	&\cos(2 \phi_a + 2 \phi_b)	\\
+ 9/8 		&3 \mod{\b1} - 3 \mod{\b2} - \mod{\c1} + \mod{\c2}	&\alpha_a	& 	& 	&\cos^3\theta_a	&1 + \cos^2\theta_b	& 	\\
\\[-3mm]\hline\\[-3mm]
- 3/8 		&\mod{\b1} - \mod{\b2} - 3 \mod{\c1} + 3 \mod{\c2}	& 	&P_z	&\cos\theta	& 	&1 + \cos^2\theta_b	& 	\\
- 3\sqrt{2/3}\big/4 	&\re{\a1}{\c1} - \re{\a2}{\c2}	& 	&P_z	&\cos\theta	&\sin2\theta_a	&\sin2\theta_b	&\cos(\phi_a + \phi_b)	\\
+ 3/4 		&\mod{\a1} - \mod{\a2}	& 	&P_z	&\cos\theta	&(1 + 3 \cos^2\theta_a)	&\sin^2\theta_b	& 	\\
+ 3\sqrt3\big/4 	&\re{\b1}{\c2} - \re{\b2}{\c1}	& 	&P_z	&\cos\theta	&\sin^2\theta_a	&\sin^2\theta_b	&\cos(2 \phi_a + 2 \phi_b)	\\
- 9/8 		&\mod{\b1} - \mod{\b2} + \mod{\c1} - \mod{\c2}	& 	&P_z	&\cos\theta	&\cos^2\theta_a	&1 + \cos^2\theta_b	& 	\\
+ 3\sqrt3\big/4 	&\re{\b1}{\c1} - \re{\b2}{\c2}	& 	&P_z	&\sin\theta	&\sin2\theta_a	&1 + \cos^2\theta_b	&\cos\phi_a	\\
- 3\big/4\sqrt2 	&\re{\a1}{\b1} - \re{\a2}{\b2}	& 	&P_z	&\sin\theta	&(1 + 3 \cos^2\theta_a)	&\sin2\theta_b	&\cos\phi_b	\\
- 3\sqrt{2/3}\big/4 	&\re{\a1}{\c2} - \re{\a2}{\c1}	& 	&P_z	&\sin\theta	&\sin^2\theta_a	&\sin2\theta_b	&\cos(2 \phi_a + \phi_b)	\\
+ 3/8 		&5 \mod{\b1} + 5 \mod{\b2} + 3 \mod{\c1} + 3 \mod{\c2}	&\alpha_a	&P_z	&\cos\theta	&\cos\theta_a	&1 + \cos^2\theta_b	& 	\\
- 3/4 		&\mod{\a1} + \mod{\a2}	&\alpha_a	&P_z	&\cos\theta	&(5 - 9 \cos^2\theta_a) \cos\theta_a	&\sin^2\theta_b	& 	\\
+ 3\sqrt{2/3}\big/4 	&\re{\a1}{\c1} + \re{\a2}{\c2}	&\alpha_a	&P_z	&\cos\theta	&(1 - 3 \cos^2\theta_a) \sin\theta_a	&\sin2\theta_b	&\cos(\phi_a + \phi_b)	\\
+ 3\big/4\sqrt2 	&\re{\a1}{\b2} + \re{\a2}{\b1}	&\alpha_a	&P_z	&\cos\theta	&(1 - 9 \cos^2\theta_a) \sin\theta_a	&\sin2\theta_b	&\cos(\phi_a + \phi_b)	\\
- 9\sqrt3\big/4 	&\re{\b1}{\c2} + \re{\b2}{\c1}	&\alpha_a	&P_z	&\cos\theta	&\sin^2\theta_a \cos\theta_a	&\sin^2\theta_b	&\cos(2 \phi_a + 2 \phi_b)	\\
- 9/8 		&3 \mod{\b1} + 3 \mod{\b2} + \mod{\c1} + \mod{\c2}	&\alpha_a	&P_z	&\cos\theta	&\cos^3\theta_a	&1 + \cos^2\theta_b	& 	\\
+ 3\big/4\sqrt2 	&\re{\a1}{\b1} + \re{\a2}{\b2}	&\alpha_a	&P_z	&\sin\theta	&(5 - 9 \cos^2\theta_a) \cos\theta_a	&\sin2\theta_b	&\cos\phi_b	\\
- 3\sqrt3\big/4 	&\re{\b1}{\c1} + \re{\b2}{\c2}	&\alpha_a	&P_z	&\sin\theta	&(1 - 3 \cos^2\theta_a) \sin\theta_a	&1 + \cos^2\theta_b	&\cos\phi_a	\\
+ 3/2 		&\re{\a1}{\a2}	&\alpha_a	&P_z	&\sin\theta	&(1 - 9 \cos^2\theta_a) \sin\theta_a	&\sin^2\theta_b	&\cos\phi_a	\\
+ 3/4 		&\re{\b1}{\b2}	&\alpha_a	&P_z	&\sin\theta	&(1 - 9 \cos^2\theta_a) \sin\theta_a	&\sin^2\theta_b	&\cos(\phi_a + 2 \phi_b)	\\
+ 9/4 		&\re{\c1}{\c2}	&\alpha_a	&P_z	&\sin\theta	&\sin^3\theta_a	&\sin^2\theta_b	&\cos(3 \phi_a + 2 \phi_b)	\\
+ 9\sqrt{2/3}\big/4 	&\re{\a1}{\c2} + \re{\a2}{\c1}	&\alpha_a	&P_z	&\sin\theta	&\sin^2\theta_a \cos\theta_a	&\sin2\theta_b	&\cos(2 \phi_a + \phi_b)	\\
\\[-3mm]\hline\\[-3mm]
+ 3\sqrt{2/3}\big/4 	&\im{\a1}{\c1} + \im{\a2}{\c2}	& 	&P_z	&\cos\theta	&\sin2\theta_a	&\sin2\theta_b	&\sin(\phi_a + \phi_b)	\\
+ 3\sqrt3\big/4 	&\im{\b1}{\c2} + \im{\b2}{\c1}	& 	&P_z	&\cos\theta	&\sin^2\theta_a	&\sin^2\theta_b	&\sin(2 \phi_a + 2 \phi_b)	\\
- 3\sqrt3\big/4 	&\im{\b1}{\c1} + \im{\b2}{\c2}	& 	&P_z	&\sin\theta	&\sin2\theta_a	&1 + \cos^2\theta_b	&\sin\phi_a	\\
+ 3\big/4\sqrt2 	&\im{\a1}{\b1} + \im{\a2}{\b2}	& 	&P_z	&\sin\theta	&(1 + 3 \cos^2\theta_a)	&\sin2\theta_b	&\sin\phi_b	\\
- 3\sqrt{2/3}\big/4 	&\im{\a1}{\c2} + \im{\a2}{\c1}	& 	&P_z	&\sin\theta	&\sin^2\theta_a	&\sin2\theta_b	&\sin(2 \phi_a + \phi_b)	\\
- 3\sqrt{2/3}\big/4 	&\im{\a1}{\c1} - \im{\a2}{\c2}	&\alpha_a	&P_z	&\cos\theta	&(1 - 3 \cos^2\theta_a) \sin\theta_a	&\sin2\theta_b	&\sin(\phi_a + \phi_b)	\\
+ 3\big/4\sqrt2 	&\im{\a1}{\b2} - \im{\a2}{\b1}	&\alpha_a	&P_z	&\cos\theta	&(1 - 9 \cos^2\theta_a) \sin\theta_a	&\sin2\theta_b	&\sin(\phi_a + \phi_b)	\\
- 9\sqrt3\big/4 	&\im{\b1}{\c2} - \im{\b2}{\c1}	&\alpha_a	&P_z	&\cos\theta	&\sin^2\theta_a \cos\theta_a	&\sin^2\theta_b	&\sin(2 \phi_a + 2 \phi_b)	\\
- 3\big/4\sqrt2 	&\im{\a1}{\b1} - \im{\a2}{\b2}	&\alpha_a	&P_z	&\sin\theta	&(5 - 9 \cos^2\theta_a) \cos\theta_a	&\sin2\theta_b	&\sin\phi_b	\\
+ 3\sqrt3\big/4 	&\im{\b1}{\c1} - \im{\b2}{\c2}	&\alpha_a	&P_z	&\sin\theta	&(1 - 3 \cos^2\theta_a) \sin\theta_a	&1 + \cos^2\theta_b	&\sin\phi_a	\\
+ 3/2 		&\im{\a1}{\a2}	&\alpha_a	&P_z	&\sin\theta	&(1 - 9 \cos^2\theta_a) \sin\theta_a	&\sin^2\theta_b	&\sin\phi_a	\\
+ 3/4 		&\im{\b1}{\b2}	&\alpha_a	&P_z	&\sin\theta	&(1 - 9 \cos^2\theta_a) \sin\theta_a	&\sin^2\theta_b	&\sin(\phi_a + 2 \phi_b)	\\
+ 9/4 		&\im{\c1}{\c2}	&\alpha_a	&P_z	&\sin\theta	&\sin^3\theta_a	&\sin^2\theta_b	&\sin(3 \phi_a + 2 \phi_b)	\\
+ 9\sqrt{2/3}\big/4 	&\im{\a1}{\c2} - \im{\a2}{\c1}	&\alpha_a	&P_z	&\sin\theta	&\sin^2\theta_a \cos\theta_a	&\sin2\theta_b	&\sin(2 \phi_a + \phi_b)	\\
\\[-3mm]\hline\\[-3mm]
+ 3\sqrt{2/3}\big/4 	&\im{\a1}{\c1} - \im{\a2}{\c2}	& 	& 	& 	&\sin2\theta_a	&\sin2\theta_b	&\sin(\phi_a + \phi_b)	\\
- 3\sqrt3\big/4 	&\im{\b1}{\c2} - \im{\b2}{\c1}	& 	& 	& 	&\sin^2\theta_a	&\sin^2\theta_b	&\sin(2 \phi_a + 2 \phi_b)	\\
- 3\sqrt{2/3}\big/4 	&\im{\a1}{\c1} + \im{\a2}{\c2}	&\alpha_a	& 	& 	&(1 - 3 \cos^2\theta_a) \sin\theta_a	&\sin2\theta_b	&\sin(\phi_a + \phi_b)	\\
+ 3\big/4\sqrt2 	&\im{\a1}{\b2} + \im{\a2}{\b1}	&\alpha_a	& 	& 	&(1 - 9 \cos^2\theta_a) \sin\theta_a	&\sin2\theta_b	&\sin(\phi_a + \phi_b)	\\
+ 9\sqrt3\big/4 	&\im{\b1}{\c2} + \im{\b2}{\c1}	&\alpha_a	& 	& 	&\sin^2\theta_a \cos\theta_a	&\sin^2\theta_b	&\sin(2 \phi_a + 2 \phi_b)	\\

	\\[-3mm]\hline
	\end{array}%
}%
}%
\caption{Same as \autoref{tab:proc1}, for the $0^{1/2}\to a^{3/2}b^{1}\to
         1^{1/2}\,2^{0}\:3^{1/2}\,4^{1/2}$ process where particle $a$ has spin
         $3/2$ instead of $1/2$.}
\label{tab:proc5}
\vspace*{-2cm}
\end{table}

\begin{table}\center
\ensuremath{%
	\begin{array}{@{}  *{7}{c@{\quad}}  c@{}}
	\hline\\[-3mm]
	+ 3 		&\mod{\a1} + \mod{\a2}	& 	& 	& 	& 	&\cos^2\theta_b	& 	\\
+ 3/2 		&\mod{\b1} + \mod{\b2}	& 	& 	& 	& 	&\sin^2\theta_b	& 	\\
+ 3 		&\mod{\a1} - \mod{\a2}	&\alpha_a	& 	& 	&\cos\theta_a	&\cos^2\theta_b	& 	\\
+ 3/2 		&\mod{\b1} - \mod{\b2}	&\alpha_a	& 	& 	&\cos\theta_a	&\sin^2\theta_b	& 	\\
+ 3 	/\sqrt2 	&\re{\a1}{\b2} - \re{\a2}{\b1}	&\alpha_a	& 	& 	&\sin\theta_a	&\sin2\theta_b	&\cos(\phi_a + \phi_b)	\\
\\[-3mm]\hline\\[-3mm]
+ 3 		&\mod{\a1} - \mod{\a2}	& 	&P_z	&\cos\theta	& 	&\cos^2\theta_b	& 	\\
- 3/2 		&\mod{\b1} - \mod{\b2}	& 	&P_z	&\cos\theta	& 	&\sin^2\theta_b	& 	\\
+ 3 	/\sqrt2 	&\re{\a1}{\b1} - \re{\a2}{\b2}	& 	&P_z	&\sin\theta	& 	&\sin2\theta_b	&\cos\phi_b	\\
+ 3 		&\mod{\a1} + \mod{\a2}	&\alpha_a	&P_z	&\cos\theta	&\cos\theta_a	&\cos^2\theta_b	& 	\\
- 3/2 		&\mod{\b1} + \mod{\b2}	&\alpha_a	&P_z	&\cos\theta	&\cos\theta_a	&\sin^2\theta_b	& 	\\
+ 3 	/\sqrt2 	&\re{\a1}{\b2} + \re{\a2}{\b1}	&\alpha_a	&P_z	&\cos\theta	&\sin\theta_a	&\sin2\theta_b	&\cos(\phi_a + \phi_b)	\\
+ 3 	/\sqrt2 	&\re{\a1}{\b1} + \re{\a2}{\b2}	&\alpha_a	&P_z	&\sin\theta	&\cos\theta_a	&\sin2\theta_b	&\cos\phi_b	\\
- 6 		&\re{\a1}{\a2}	&\alpha_a	&P_z	&\sin\theta	&\sin\theta_a	&\cos^2\theta_b	&\cos\phi_a	\\
+ 3 		&\re{\b1}{\b2}	&\alpha_a	&P_z	&\sin\theta	&\sin\theta_a	&\sin^2\theta_b	&\cos(\phi_a + 2 \phi_b)	\\
\\[-3mm]\hline\\[-3mm]
- 3 	/\sqrt2 	&\im{\a1}{\b1} + \im{\a2}{\b2}	& 	&P_z	&\sin\theta	& 	&\sin2\theta_b	&\sin\phi_b	\\
+ 3 	/\sqrt2 	&\im{\a1}{\b2} - \im{\a2}{\b1}	&\alpha_a	&P_z	&\cos\theta	&\sin\theta_a	&\sin2\theta_b	&\sin(\phi_a + \phi_b)	\\
- 3 	/\sqrt2 	&\im{\a1}{\b1} - \im{\a2}{\b2}	&\alpha_a	&P_z	&\sin\theta	&\cos\theta_a	&\sin2\theta_b	&\sin\phi_b	\\
- 6 		&\im{\a1}{\a2}	&\alpha_a	&P_z	&\sin\theta	&\sin\theta_a	&\cos^2\theta_b	&\sin\phi_a	\\
+ 3 		&\im{\b1}{\b2}	&\alpha_a	&P_z	&\sin\theta	&\sin\theta_a	&\sin^2\theta_b	&\sin(\phi_a + 2 \phi_b)	\\
\\[-3mm]\hline\\[-3mm]
+ 3 	/\sqrt2 	&\im{\a1}{\b2} + \im{\a2}{\b1}	&\alpha_a	& 	& 	&\sin\theta_a	&\sin2\theta_b	&\sin(\phi_a + \phi_b)	\\

	\\[-3mm]\hline
	\end{array}%
}%
\caption{Various contributions to the angular distribution of the $0^{1/2}\to
         a^{1/2}b^{1}\to 1^{1/2}\,2^{0}\:3^{0}\,4^{0}$ process, with conventions
         and assumptions specified in the text. Each line corresponds to a term
         of different angular dependence (most of them being independent). The
         separation in columns is only meant to ease the comparison between the
         various factors appearing in each term. The four blocks distinguish
         terms whose combinations of angular and polarisation dependence have
         different parity and motion reversal transformation properties. They
         are respectively \P-even--\T-even, \P-odd--\T-even, \P-even--\T-odd,
         and \P-odd--\T-odd.}
\label{tab:proc2}
\end{table}

\begin{table}\center
\vspace{-1cm}
\scalebox{.85}{%
\ensuremath{%
	\begin{array}{@{}  *{7}{c@{\;\;}}  c@{}}
	\hline\\[-3mm]
	+ 3/4 		&\mod{\b1} + \mod{\b2} + 3 \mod{\c1} + 3 \mod{\c2}	& 	& 	& 	& 	&\sin^2\theta_b	& 	\\
+ 3\sqrt{2/3}\big/2 	&\re{\a1}{\c1} + \re{\a2}{\c2}	& 	& 	& 	&\sin2\theta_a	&\sin2\theta_b	&\cos(\phi_a + \phi_b)	\\
+ 3/2 		&\mod{\a1} + \mod{\a2}	& 	& 	& 	&(1 + 3 \cos^2\theta_a)	&\cos^2\theta_b	& 	\\
+ 3\sqrt3\big/2 	&\re{\b1}{\c2} + \re{\b2}{\c1}	& 	& 	& 	&\sin^2\theta_a	&\sin^2\theta_b	&\cos(2 \phi_a + 2 \phi_b)	\\
+ 9/4 		&\mod{\b1} + \mod{\b2} - \mod{\c1} - \mod{\c2}	& 	& 	& 	&\cos^2\theta_a	&\sin^2\theta_b	& 	\\
- 3/4 		&5 \mod{\b1} - 5 \mod{\b2} - 3 \mod{\c1} + 3 \mod{\c2}	&\alpha_a	& 	& 	&\cos\theta_a	&\sin^2\theta_b	& 	\\
- 3/2 		&\mod{\a1} - \mod{\a2}	&\alpha_a	& 	& 	&(5 - 9 \cos^2\theta_a) \cos\theta_a	&\cos^2\theta_b	& 	\\
- 3\sqrt{2/3}\big/2 	&\re{\a1}{\c1} - \re{\a2}{\c2}	&\alpha_a	& 	& 	&(1 - 3 \cos^2\theta_a) \sin\theta_a	&\sin2\theta_b	&\cos(\phi_a + \phi_b)	\\
- 3\big/2\sqrt2 	&\re{\a1}{\b2} - \re{\a2}{\b1}	&\alpha_a	& 	& 	&(1 - 9 \cos^2\theta_a) \sin\theta_a	&\sin2\theta_b	&\cos(\phi_a + \phi_b)	\\
- 9\sqrt3\big/2 	&\re{\b1}{\c2} - \re{\b2}{\c1}	&\alpha_a	& 	& 	&\sin^2\theta_a \cos\theta_a	&\sin^2\theta_b	&\cos(2 \phi_a + 2 \phi_b)	\\
+ 9/4 		&3 \mod{\b1} - 3 \mod{\b2} - \mod{\c1} + \mod{\c2}	&\alpha_a	& 	& 	&\cos^3\theta_a	&\sin^2\theta_b	& 	\\
\\[-3mm]\hline\\[-3mm]
- 3/4 		&\mod{\b1} - \mod{\b2} - 3 \mod{\c1} + 3 \mod{\c2}	& 	&P_z	&\cos\theta	& 	&\sin^2\theta_b	& 	\\
+ 3\sqrt{2/3}\big/2 	&\re{\a1}{\c1} - \re{\a2}{\c2}	& 	&P_z	&\cos\theta	&\sin2\theta_a	&\sin2\theta_b	&\cos(\phi_a + \phi_b)	\\
+ 3/2 		&\mod{\a1} - \mod{\a2}	& 	&P_z	&\cos\theta	&(1 + 3 \cos^2\theta_a)	&\cos^2\theta_b	& 	\\
- 3\sqrt3\big/2 	&\re{\b1}{\c2} - \re{\b2}{\c1}	& 	&P_z	&\cos\theta	&\sin^2\theta_a	&\sin^2\theta_b	&\cos(2 \phi_a + 2 \phi_b)	\\
- 9/4 		&\mod{\b1} - \mod{\b2} + \mod{\c1} - \mod{\c2}	& 	&P_z	&\cos\theta	&\cos^2\theta_a	&\sin^2\theta_b	& 	\\
+ 3\sqrt3\big/2 	&\re{\b1}{\c1} - \re{\b2}{\c2}	& 	&P_z	&\sin\theta	&\sin2\theta_a	&\sin^2\theta_b	&\cos\phi_a	\\
+ 3\big/2\sqrt2 	&\re{\a1}{\b1} - \re{\a2}{\b2}	& 	&P_z	&\sin\theta	&(1 + 3 \cos^2\theta_a)	&\sin2\theta_b	&\cos\phi_b	\\
+ 3\sqrt{2/3}\big/2 	&\re{\a1}{\c2} - \re{\a2}{\c1}	& 	&P_z	&\sin\theta	&\sin^2\theta_a	&\sin2\theta_b	&\cos(2 \phi_a + \phi_b)	\\
+ 3/4 		&5 \mod{\b1} + 5 \mod{\b2} + 3 \mod{\c1} + 3 \mod{\c2}	&\alpha_a	&P_z	&\cos\theta	&\cos\theta_a	&\sin^2\theta_b	& 	\\
- 3/2 		&\mod{\a1} + \mod{\a2}	&\alpha_a	&P_z	&\cos\theta	&(5 - 9 \cos^2\theta_a) \cos\theta_a	&\cos^2\theta_b	& 	\\
- 3\sqrt{2/3}\big/2 	&\re{\a1}{\c1} + \re{\a2}{\c2}	&\alpha_a	&P_z	&\cos\theta	&(1 - 3 \cos^2\theta_a) \sin\theta_a	&\sin2\theta_b	&\cos(\phi_a + \phi_b)	\\
- 3\big/2\sqrt2 	&\re{\a1}{\b2} + \re{\a2}{\b1}	&\alpha_a	&P_z	&\cos\theta	&(1 - 9 \cos^2\theta_a) \sin\theta_a	&\sin2\theta_b	&\cos(\phi_a + \phi_b)	\\
+ 9\sqrt3\big/2 	&\re{\b1}{\c2} + \re{\b2}{\c1}	&\alpha_a	&P_z	&\cos\theta	&\sin^2\theta_a \cos\theta_a	&\sin^2\theta_b	&\cos(2 \phi_a + 2 \phi_b)	\\
- 9/4 		&3 \mod{\b1} + 3 \mod{\b2} + \mod{\c1} + \mod{\c2}	&\alpha_a	&P_z	&\cos\theta	&\cos^3\theta_a	&\sin^2\theta_b	& 	\\
- 3\big/2\sqrt2 	&\re{\a1}{\b1} + \re{\a2}{\b2}	&\alpha_a	&P_z	&\sin\theta	&(5 - 9 \cos^2\theta_a) \cos\theta_a	&\sin2\theta_b	&\cos\phi_b	\\
- 3\sqrt3\big/2 	&\re{\b1}{\c1} + \re{\b2}{\c2}	&\alpha_a	&P_z	&\sin\theta	&(1 - 3 \cos^2\theta_a) \sin\theta_a	&\sin^2\theta_b	&\cos\phi_a	\\
+ 3 		&\re{\a1}{\a2}	&\alpha_a	&P_z	&\sin\theta	&(1 - 9 \cos^2\theta_a) \sin\theta_a	&\cos^2\theta_b	&\cos\phi_a	\\
- 3/2 		&\re{\b1}{\b2}	&\alpha_a	&P_z	&\sin\theta	&(1 - 9 \cos^2\theta_a) \sin\theta_a	&\sin^2\theta_b	&\cos(\phi_a + 2 \phi_b)	\\
- 9/2 		&\re{\c1}{\c2}	&\alpha_a	&P_z	&\sin\theta	&\sin^3\theta_a	&\sin^2\theta_b	&\cos(3 \phi_a + 2 \phi_b)	\\
- 9\sqrt{2/3}\big/2 	&\re{\a1}{\c2} + \re{\a2}{\c1}	&\alpha_a	&P_z	&\sin\theta	&\sin^2\theta_a \cos\theta_a	&\sin2\theta_b	&\cos(2 \phi_a + \phi_b)	\\
\\[-3mm]\hline\\[-3mm]
- 3\sqrt{2/3}\big/2 	&\im{\a1}{\c1} + \im{\a2}{\c2}	& 	&P_z	&\cos\theta	&\sin2\theta_a	&\sin2\theta_b	&\sin(\phi_a + \phi_b)	\\
- 3\sqrt3\big/2 	&\im{\b1}{\c2} + \im{\b2}{\c1}	& 	&P_z	&\cos\theta	&\sin^2\theta_a	&\sin^2\theta_b	&\sin(2 \phi_a + 2 \phi_b)	\\
- 3\sqrt3\big/2 	&\im{\b1}{\c1} + \im{\b2}{\c2}	& 	&P_z	&\sin\theta	&\sin2\theta_a	&\sin^2\theta_b	&\sin\phi_a	\\
- 3\big/2\sqrt2 	&\im{\a1}{\b1} + \im{\a2}{\b2}	& 	&P_z	&\sin\theta	&(1 + 3 \cos^2\theta_a)	&\sin2\theta_b	&\sin\phi_b	\\
+ 3\sqrt{2/3}\big/2 	&\im{\a1}{\c2} + \im{\a2}{\c1}	& 	&P_z	&\sin\theta	&\sin^2\theta_a	&\sin2\theta_b	&\sin(2 \phi_a + \phi_b)	\\
+ 3\sqrt{2/3}\big/2 	&\im{\a1}{\c1} - \im{\a2}{\c2}	&\alpha_a	&P_z	&\cos\theta	&(1 - 3 \cos^2\theta_a) \sin\theta_a	&\sin2\theta_b	&\sin(\phi_a + \phi_b)	\\
- 3\big/2\sqrt2 	&\im{\a1}{\b2} - \im{\a2}{\b1}	&\alpha_a	&P_z	&\cos\theta	&(1 - 9 \cos^2\theta_a) \sin\theta_a	&\sin2\theta_b	&\sin(\phi_a + \phi_b)	\\
+ 9\sqrt3\big/2 	&\im{\b1}{\c2} - \im{\b2}{\c1}	&\alpha_a	&P_z	&\cos\theta	&\sin^2\theta_a \cos\theta_a	&\sin^2\theta_b	&\sin(2 \phi_a + 2 \phi_b)	\\
+ 3\big/2\sqrt2 	&\im{\a1}{\b1} - \im{\a2}{\b2}	&\alpha_a	&P_z	&\sin\theta	&(5 - 9 \cos^2\theta_a) \cos\theta_a	&\sin2\theta_b	&\sin\phi_b	\\
+ 3\sqrt3\big/2 	&\im{\b1}{\c1} - \im{\b2}{\c2}	&\alpha_a	&P_z	&\sin\theta	&(1 - 3 \cos^2\theta_a) \sin\theta_a	&\sin^2\theta_b	&\sin\phi_a	\\
+ 3 		&\im{\a1}{\a2}	&\alpha_a	&P_z	&\sin\theta	&(1 - 9 \cos^2\theta_a) \sin\theta_a	&\cos^2\theta_b	&\sin\phi_a	\\
- 3/2 		&\im{\b1}{\b2}	&\alpha_a	&P_z	&\sin\theta	&(1 - 9 \cos^2\theta_a) \sin\theta_a	&\sin^2\theta_b	&\sin(\phi_a + 2 \phi_b)	\\
- 9/2 		&\im{\c1}{\c2}	&\alpha_a	&P_z	&\sin\theta	&\sin^3\theta_a	&\sin^2\theta_b	&\sin(3 \phi_a + 2 \phi_b)	\\
- 9\sqrt{2/3}\big/2 	&\im{\a1}{\c2} - \im{\a2}{\c1}	&\alpha_a	&P_z	&\sin\theta	&\sin^2\theta_a \cos\theta_a	&\sin2\theta_b	&\sin(2 \phi_a + \phi_b)	\\
\\[-3mm]\hline\\[-3mm]
- 3\sqrt{2/3}\big/2 	&\im{\a1}{\c1} - \im{\a2}{\c2}	& 	& 	& 	&\sin2\theta_a	&\sin2\theta_b	&\sin(\phi_a + \phi_b)	\\
+ 3\sqrt3\big/2 	&\im{\b1}{\c2} - \im{\b2}{\c1}	& 	& 	& 	&\sin^2\theta_a	&\sin^2\theta_b	&\sin(2 \phi_a + 2 \phi_b)	\\
+ 3\sqrt{2/3}\big/2 	&\im{\a1}{\c1} + \im{\a2}{\c2}	&\alpha_a	& 	& 	&(1 - 3 \cos^2\theta_a) \sin\theta_a	&\sin2\theta_b	&\sin(\phi_a + \phi_b)	\\
- 3\big/2\sqrt2 	&\im{\a1}{\b2} + \im{\a2}{\b1}	&\alpha_a	& 	& 	&(1 - 9 \cos^2\theta_a) \sin\theta_a	&\sin2\theta_b	&\sin(\phi_a + \phi_b)	\\
- 9\sqrt3\big/2 	&\im{\b1}{\c2} + \im{\b2}{\c1}	&\alpha_a	& 	& 	&\sin^2\theta_a \cos\theta_a	&\sin^2\theta_b	&\sin(2 \phi_a + 2 \phi_b)	\\

	\\[-3mm]\hline
	\end{array}%
}%
}%
\caption{Same as \autoref{tab:proc2}, for the $0^{1/2}\to a^{3/2}b^{1}\to
         1^{1/2}\,2^{0}\:3^{0}\,4^{0}$ process, where particle $a$ has spin
         $3/2$ instead of $1/2$.}
\label{tab:proc6}
\vspace*{-2.5cm}
\end{table}

\begin{table}\center
\ensuremath{%
	\begin{array}{@{}  *{6}{c@{\quad}}  c@{}}
	\hline\\[-3mm]
	+ 		&\mod{\b1} + \mod{\b2}	& 	& 	& 	& 	& 	\\
+ 		&\mod{\b1} - \mod{\b2}	&\alpha_a	& 	& 	&\cos\theta_a	& 	\\
\\[-3mm]\hline\\[-3mm]
- 		&\mod{\b1} - \mod{\b2}	& 	&P_z	&\cos\theta	& 	& 	\\
- 		&\mod{\b1} + \mod{\b2}	&\alpha_a	&P_z	&\cos\theta	&\cos\theta_a	& 	\\

	\\[-3mm]\hline
	\end{array}
}
\caption{Contributions to the angular distribution of the three-body $0^{1/2}\to
         a^{1/2}b^{1}\to 1^{1/2}\,2^{0}\:b^{1}$ decay, under the conventions and
         assumptions specified in the text. The four blocks distinguish terms
         whose combinations of angular and polarisation dependence are
         respectively \P-even--\T-even and \P-odd--\T-even. A third block, which
         receives no contributions here, includes \P-even--\T-odd decay
         terms in the subsequent tables of this series.}
\label{tab:proc_3body_1}
\end{table}

\begin{table}\center
\ensuremath{%
	\begin{array}{@{}  *{6}{c@{\quad}}  c@{}}
	\hline\\[-3mm]
	+ 1/2 		&\mod{\b1} + \mod{\b2}	& 	& 	& 	&(1 + 3 \cos^2\theta_a)	& 	\\
+ 3/2 		&\mod{\c1} + \mod{\c2}	& 	& 	& 	&\sin^2\theta_a	& 	\\
- 1/2 		&\mod{\b1} - \mod{\b2}	&\alpha_a	& 	& 	&(5 - 9 \cos^2\theta_a) \cos\theta_a	& 	\\
+ 3/4 		&\mod{\c1} - \mod{\c2}	&\alpha_a	& 	& 	&\sin\theta_a \sin2\theta_a	& 	\\
\\[-3mm]\hline\\[-3mm]
+ 	\sqrt3 	&\re{\b1}{\c1} - \re{\b2}{\c2}	& 	&P_z	&\sin\theta	&\sin2\theta_a	&\cos\phi_a	\\
- 1/2 		&\mod{\b1} - \mod{\b2}	& 	&P_z	&\cos\theta	&(1 + 3 \cos^2\theta_a)	& 	\\
+ 3/2 		&\mod{\c1} - \mod{\c2}	& 	&P_z	&\cos\theta	&\sin^2\theta_a	& 	\\
- 	\sqrt3 	&\re{\b1}{\c1} + \re{\b2}{\c2}	&\alpha_a	&P_z	&\sin\theta	&(1 - 3 \cos^2\theta_a) \sin\theta_a	&\cos\phi_a	\\
+ 1/2 		&\mod{\b1} + \mod{\b2}	&\alpha_a	&P_z	&\cos\theta	&(5 - 9 \cos^2\theta_a) \cos\theta_a	& 	\\
+ 3/4 		&\mod{\c1} + \mod{\c2}	&\alpha_a	&P_z	&\cos\theta	&\sin\theta_a \sin2\theta_a	& 	\\
\\[-3mm]\hline\\[-3mm]
- 	\sqrt3 	&\im{\b1}{\c1} + \im{\b2}{\c2}	& 	&P_z	&\sin\theta	&\sin2\theta_a	&\sin\phi_a	\\
+ 	\sqrt3 	&\im{\b1}{\c1} - \im{\b2}{\c2}	&\alpha_a	&P_z	&\sin\theta	&(1 - 3 \cos^2\theta_a) \sin\theta_a	&\sin\phi_a	\\

	\\[-3mm]\hline
	\end{array}
}
\caption{Same as \autoref{tab:proc_3body_1}, for the $0^{1/2}\to a^{3/2}b^{1}\to
         1^{1/2}\,2^{0}\:b^{1}$ process, where particle $a$ has spin $3/2$
         instead of $1/2$.}
\label{tab:proc_3body_2}
\end{table}

\begin{table}\center
\ensuremath{%
	\begin{array}{@{}  *{6}{c@{\quad}}  c@{}}
	\hline\\[-3mm]
	+ 		&\mod{\a1} + \mod{\a2}	& 	& 	& 	& 	& 	\\
+ 		&\mod{\a1} - \mod{\a2}	&\alpha_a	& 	& 	&\cos\theta_a	& 	\\
\\[-3mm]\hline\\[-3mm]
+ 		&\mod{\a1} - \mod{\a2}	& 	&P_z	&\cos\theta	& 	& 	\\
- 2 		&\re{\a1}{\a2}	&\alpha_a	&P_z	&\sin\theta	&\sin\theta_a	&\cos\phi_a	\\
+ 		&\mod{\a1} + \mod{\a2}	&\alpha_a	&P_z	&\cos\theta	&\cos\theta_a	& 	\\
\\[-3mm]\hline\\[-3mm]
- 2 		&\im{\a1}{\a2}	&\alpha_a	&P_z	&\sin\theta	&\sin\theta_a	&\sin\phi_a	\\

	\\[-3mm]\hline
	\end{array}
}
\caption{Same as \autoref{tab:proc_3body_1}, for the $0^{1/2}\to a^{1/2}b^{0}\to
         1^{1/2}\,2^{0}\:b^{0}$ process, where particle $b$ has spin $0$ instead
         of $1$.}
\label{tab:proc_3body_3}
\end{table}

\begin{table}\center
\ensuremath{%
	\begin{array}{@{}  *{6}{c@{\quad}}  c@{}}
	\hline\\[-3mm]
	+ 1/2 		&\mod{\a1} + \mod{\a2}	& 	& 	& 	&(1 + 3 \cos^2\theta_a)	& 	\\
- 1/2 		&\mod{\a1} - \mod{\a2}	&\alpha_a	& 	& 	&(5 - 9 \cos^2\theta_a) \cos\theta_a	& 	\\
\\[-3mm]\hline\\[-3mm]
+ 1/2 		&\mod{\a1} - \mod{\a2}	& 	&P_z	&\cos\theta	&(1 + 3 \cos^2\theta_a)	& 	\\
+ 		&\re{\a1}{\a2}	&\alpha_a	&P_z	&\sin\theta	&(1 - 9 \cos^2\theta_a) \sin\theta_a	&\cos\phi_a	\\
- 1/2 		&\mod{\a1} + \mod{\a2}	&\alpha_a	&P_z	&\cos\theta	&(5 - 9 \cos^2\theta_a) \cos\theta_a	& 	\\
\\[-3mm]\hline\\[-3mm]
+ 		&\im{\a1}{\a2}	&\alpha_a	&P_z	&\sin\theta	&(1 - 9 \cos^2\theta_a) \sin\theta_a	&\sin\phi_a	\\

	\\[-3mm]\hline
	\end{array}
}
\caption{Same as \autoref{tab:proc_3body_3}, for the $0^{1/2}\to a^{3/2}b^{0}\to
         1^{1/2}\,2^{0}\:b^{0}$ process, where particle $a$ has spin $3/2$
         instead of $1/2$.}
\label{tab:proc_3body_4}
\end{table}

\section{Discrete symmetry properties}
\label{sec:prop}
To establish the parity $\P$ and motion reversal $\T$ transformation properties of the various
contributions to the differential distributions displayed in
\autoref{tab:proc1}--\ref{tab:proc_3body_4}, let us define our kinematic variables and
axes in terms of physical momenta.

\newcommand{\pI}{\ensuremath{\vec{p_A}}}
\newcommand{\pII}{\ensuremath{\vec{p_B}}}
In the restframe of particle~$0$, let us assume that the production plane is
defined by the momenta \pI\ and \pII\ of two of the particles involved. One can
then construct the $(\vec x,\vec y,\vec z)$ frame as
\begin{align*}
	\vec x	= \frac{\pI}{|\pI|}
		,\qquad
	\vec z	= \frac{\pI\times\pII}{|\pI\times\pII|}
		,\qquad
	\vec y	= \vec z\times\vec x.
\end{align*}
The $\vec z$ axis is thus a \P-even--\T-even pseudovector, while $\vec x$ and
$\vec y$ are both \P-odd--\T-odd vectors. The $(\vec{x_a},\vec{y_a}, \vec{z_a})$
system is then obtained as
\begin{equation*}
	\vec{z_a} = \frac{\vec{p_1}+\vec{p_2}}{|\vec{p_1}+\vec{p_2}|}
	,\qquad
	\vec{y_a} = \frac{\vec z\times \vec{z_a}}{|\vec z\times \vec{z_a}|}
	,\qquad
	\vec{x_a} = \frac{\vec{y_a}\times\vec{z_a}}{|\vec{y_a}\times\vec{z_a}|},
\end{equation*}
where $\vec{p_1}$ and $\vec{p_2}$ are the momenta of particles $1$ and $2$ in
particle~$0$'s restframe. It follows that both $\vec{z_a}$ and $\vec{y_a}$ are
\P-odd--\T-odd vectors while $\vec{x_a}$ is a \P-even--\T-even pseudovector.
Similar conclusions hold for $(\vec{x_b},\vec{y_b},\vec{z_b}) =
(-\vec{x_a},\vec{y_a},-\vec{z_a})$. The polar $\theta$ and azimuthal $\phi$
angles can be obtained from the equalities:
\begin{equation*}
	\cos\theta = \vec z\cdot\vec{z_a}
	,\qquad
	\sin\theta = +\sqrt{1-\cos^2\theta} 
	,\qquad
	\cos\phi = (\vec z\times \vec{z_a})\cdot \vec y
	,\qquad
	\sin\phi = -(\vec z\times \vec{z_a})\cdot \vec x
	,
\end{equation*}
which establish that $\cos\theta$ is a \P-odd--\T-odd kinematic variable, while
$\sin\theta$, $\cos\phi$, and $\sin\phi$ are \P-even--\T-even. Moreover,
defining the \P-even--\T-even pseudovectors
\begin{equation*}
	\vec{n_a}=\frac{ \vec{p_1}\times \vec{p_2} }{ |\vec{p_1}\times \vec{p_2}| }
	,\qquad
	\vec{n_b}=\frac{ \vec{p_3}\times \vec{p_4} }{ |\vec{p_3}\times \vec{p_4}| }
	,
\end{equation*}
where $\vec{p_3}$ and $\vec{p_4}$ are the momenta of particles $3$ and $4$ in
particle $0$'s restframe, one can further write
\begin{equation*}
	(0 	= \vec{n_a}\cdot \vec{z_a}
	,)\qquad
	\cos\phi_a = - \vec{n_a}\cdot \vec{y_a}
	,\qquad
	\sin\phi_a = \vec{n_a}\cdot \vec{x_a},
\end{equation*}
and similarly for $a\leftrightarrow b$. This shows that $\cos\phi_{a}$ and
$\cos\phi_{b}$ are \P-odd--\T-odd variables, while $\sin\phi_{a}$ and
$\sin\phi_{b}$ are \P-even--\T-even. With $\vec{\tilde p_1}$ and $\vec{\tilde
p_3}$, the momenta of particle $1$ and $3$, respectively measured in the
particle $a$ and $b$ restframes, one can define
\begin{equation*}
	\cos\theta_a = \vec{z_a}\cdot \frac{\vec{\tilde p_1}}{|\vec{\tilde p_1}|}
	,\qquad
	\sin\theta_a = +\sqrt{1-\cos^2\theta_a},
\end{equation*}
and similarly for $a\leftrightarrow b$ and $1\leftrightarrow 3$, demonstrating
that both $\cos\theta_{a,b}$ and $\sin\theta_{a,b}$ are \P-even--\T-even
variables. Finally, the \P-odd--\T-odd character of the $\vec z$ vector
implies that among the polarisations components
\begin{equation*}
	P_x = \langle \vec{s}\cdot \vec x \rangle
	,\qquad
	P_y = \langle \vec{s}\cdot \vec y \rangle
	,\qquad
	P_z = \langle \vec{s}\cdot \vec z \rangle
	,
\end{equation*}
$P_x$ and $P_y$ are \P-odd--\T-even, while $P_z$ is \P-even--\T-odd. Their
values are fixed by particle $0$'s production amplitudes and, in general,
depend on the production kinematics which is disregarded here.

To summarize, for our definition of frames, we have thus identified three
\P-odd--\T-odd kinematic variables:
\begin{equation*}
	\cos\theta, \quad \cos\phi_a, \quad\text{and}\quad \cos\phi_{b},
\end{equation*}
while $\cos\phi$, $\sin\phi$, $\cos\theta_{a,b}$,
$\sin\phi_{a,b}$, as well as $m_{a,b}$ which are necessary to fully specify
the final-state kinematics, are all \P-even--\T-even. The contributions to
the angular distributions we displayed in
\autoref{tab:proc1}--\ref{tab:proc_3body_4} have been grouped according to their
\P\ and \T\ transformation properties. In \autoref{tab:proc1}--\ref{tab:proc6}
relating to four-body decays, the angular distributions of the contributions in the first and
third blocks are \P-even--\T-even while that of the second and fourth ones are
\P-odd--\T-odd. The second and third blocks moreover include contributions
proportional to the \P-even--\T-odd polarisation $P_z$. As a result, the four
blocks distinguish contributions whose
combinations of angular and polarisation dependence are respectively
\P-even--\T-even, \P-odd--\T-even, \P-even--\T-odd, and \P-odd--\T-odd. In
three-body decays, there are not enough independent four-momenta to form \T-odd
$\epsilon_{\mu\nu\rho\sigma} \, p_1^\mu \, p_2^\nu \, p_3^\rho \, p_4^\sigma$
contractions. One must necessarily involve a spin four-vector. In
\autoref{tab:proc_3body_1}--\ref{tab:proc_3body_4}, all terms proportional to
imaginary parts of decay amplitude interferences therefore come proportional to
$P_z$. They thus appear in \P-even--\T-odd blocks, and there are no fourth
\P-odd--\T-odd ones. The $0^{1/2}\to a^{1/2}b^{1}\to 1^{1/2}\,2^{0}\:b^{1}$
decay relating to \autoref{tab:proc_3body_1} does moreover not contain any
term proportional to the imaginary part of decay amplitude interferences when
$b^1$ is massless (an assumption relaxed in \autoref{sec:appendix}).

\section{Asymmetries}
As mentioned in the introduction, due to the presence of a \T-odd polarisation
component $P_z$, both \T-odd and \T-even angular asymmetries can potentially
serve to access imaginary parts of decay amplitude interferences.

\subsection{\T-odd angular asymmetries}

In the spirit of \rcite{Durieux:2015zwa}, \T-odd--\CP-odd angular
asymmetries could be constructed systematically as
\begin{equation*}
	\mathcal{A}^{jkl}_{mno}\equiv
	\int\!\!\text{d}\Omega
	\left(
		\frac{1}{\Gamma}\frac{\text{d}\Gamma}{\text{d}\Omega}-
		\frac{1}{\bar\Gamma}\frac{\text{d}\bar\Gamma}{\text{d}\Omega}
	\right)
	\sign\left\{
		f_j(\cos\theta)		\:
		f_k(\cos\theta_a)	\:
		f_l(\cos\theta_b)	\:
		\sin\!\left(m\phi_a+n\phi_b+o\frac{\pi}{2}\right) \right\}
\end{equation*}
for $f_0(x)=1$, $f_1(x)=x$, $f_2(x)=3x^2-1$, etc. which could be chosen as
Legendre polynomials and various $j,k,l,m,n,o$ combinations of integers
satisfying $j+m+n+o\in 2\mathbb{Z}$ with $o\in\{0,1\}$. Contributions not
explicitly listed in the various tables of this paper could appear in the
interferences of amplitudes featuring $a$ and $b$ intermediate states of various
spins or different topologies. It was also noted in \rcite{Durieux:2015zwa} that
distinguishing regions in the $m_{a,b}$ invariant mass integration could be
useful when resonances are identified, and that pairings of final-state particles
different from the $a=(12)$, $b=(34)$ ones could increase the sensitivity to
phase differences between amplitudes of different resonance structures.
Understanding the origin of the various angular distribution components is
however required to determine whether a symmetry violation observed arises from
the decay or production, given the lack of decoupling between the two parts of
the process.

This understanding we gained in the previous section allows us to be more
specific. In both four-body processes featuring a spin-$1/2$ intermediate
resonance $a$, under the assumptions stated, there is actually one single \T-odd
angular distribution that provides access to \CP-odd phase differences between
decay amplitudes, without requiring \CP-even phase between neither decay nor
production amplitudes. The corresponding term is displayed in the fourth blocks
of \hyperref[tab:proc1]{Tables~\ref{tab:proc1}} and \ref{tab:proc2}. By relying
on the
\begin{equation*}
	\frac{1}{\Gamma}
	\int\!\!\text{d}\Omega
	\frac{\text{d}\Gamma}{\text{d}\Omega}
	\sign\{ \cos\theta_b \sin(\phi_a+\phi_b) \}
\end{equation*}
asymmetry, or on the analogue moment, one gets access to the
$(\im{\a1}{\b2}+\im{\a2}{\b1})\alpha_a$ combination of decay amplitudes. Then
combining the \CP-conjugate $0\to 1\,2\,3\,4$ and $\bar0\to
\bar1\,\bar2\,\bar3\,\bar4$ processes to form $\mathcal{A}^{001}_{110}$ yields
sensitivity to small differences in the \CP-odd phases between the $\a1\b2$ or
$\a2\b1$ amplitudes. It is maximal when they have identical \CP-even phases.
In both four-body processes featuring a spin $3/2$ intermediate resonance $a$,
one could moreover employ on the
\begin{align*}
\mathcal{A}^{011}_{110} = 
	\int\!\!\text{d}\Omega
	\left(
		\frac{1}{\Gamma}\frac{\text{d}\Gamma}{\text{d}\Omega}-
		\frac{1}{\bar\Gamma}\frac{\text{d}\bar\Gamma}{\text{d}\Omega}
	\right)
	&\sign\{ \cos\theta_a \cos\theta_b \sin(\phi_a+\phi_b) \}
	\\
\mathcal{A}^{000}_{220} = 
	\int\!\!\text{d}\Omega
	\left(
		\frac{1}{\Gamma}\frac{\text{d}\Gamma}{\text{d}\Omega}-
		\frac{1}{\bar\Gamma}\frac{\text{d}\bar\Gamma}{\text{d}\Omega}
	\right)
	&\sign\{ \sin(2\phi_a+2\phi_b) \}
	\\[2mm]
\mathcal{A}^{021}_{110} = 
	\int\!\!\text{d}\Omega
	\left(
		\frac{1}{\Gamma}\frac{\text{d}\Gamma}{\text{d}\Omega}-
		\frac{1}{\bar\Gamma}\frac{\text{d}\bar\Gamma}{\text{d}\Omega}
	\right)
	&\sign\{ (3\cos^2\theta_a-1) \cos\theta_b \sin(\phi_a+\phi_b) \}
	\\
\mathcal{A}^{010}_{220} = 
	\int\!\!\text{d}\Omega
	\left(
		\frac{1}{\Gamma}\frac{\text{d}\Gamma}{\text{d}\Omega}-
		\frac{1}{\bar\Gamma}\frac{\text{d}\bar\Gamma}{\text{d}\Omega}
	\right)
	&\sign\{ \cos\theta_a \sin(2\phi_a+2\phi_b) \}
\end{align*}
asymmetries (see \hyperref[tab:proc5]{Tables~\ref{tab:proc5}} and
\ref{tab:proc6}). Note the latter two as well as $\mathcal{A}^{001}_{110}$ come
proportional to the asymmetry parameter $\alpha_a$ which vanishes if the
$a^{3/2}\to 1^{1/2} 2^0$ decay preserves parity, as $\Lambda(1520) \to p K$ does,
being mediated by the strong interaction. A $\:\sign\{ (1-9\cos^2\theta_a)
\cos\theta_b \sin(\phi_a + \phi_b) \}\:$ asymmetry which is not independent of
the $\:\sign\{ \cos\theta_b \sin(\phi_a + \phi_b) \}\:$ and $\:\sign\{
(3\cos^2\theta_a-1) \cos\theta_b \sin(\phi_a + \phi_b) \}\:$ ones has not been
listed.

Let us also comment on the classical $\mathcal{A}^{000}_{110}$ asymmetry, based
on $\sign\{\sin(\phi_a+\phi_b)\}$, which changes sign where the antisymmetry
contraction of the four independent external particle momenta
$\epsilon_{\mu\nu\rho\sigma}\, p_1^\mu\, p_2^\nu\, p_3^\rho\, p_4^\sigma$ does.
Its use for studying \CP\ violation in the decay of $\Lambda_b$ and $\Xi_b$
baryons was advocated in \rcite{Gronau:2015gha}. We however stress that,
contrarily to the four-body decay of spinless particles where it can play a
significant role, it vanishes in the four four-body decays considered here,
under the assumptions stated. Examining
\autoref{tab:proc1_full}--\ref{tab:proc6_full} where these assumptions are
relaxed, one realises such an asymmetry only appears proportional to the
$\alpha_b$ asymmetry parameters in the $0^{1/2}\to a^{1/2,3/2}\: b^{1}\to
1^{1/2}\, 2^{0}\: 3^{1/2}\, 4^{1/2}$ decays. Even in such processes, its
presence is therefore seen to require parity violation in the $b^1\to 3^{1/2}\,
4^{1/2}$ daughter decay (which would for instance be absent in electromagnetic
$\jpsi\to \ell^+\ell^-$ decays).

Following \rcites{Conte:2008ema}{Conte:2008ema, Leitner:2006nb,
Leitner:2006sc}, the LHCb collaboration measured the four
\begin{equation*}
	\frac{1}{\Gamma}
		\int\text{d}\Omega	\;
		\frac{\text{d}\Gamma}{\text{d}\Omega}	\;
		\sign\{\:\cos\Phi_{a,b}\;,\;\; \sin\Phi_{a,b}\:\}
\end{equation*}
asymmetries in the $\Lambda_b\to\Lambda\,\varphi\to p\,\pi\,K^+K^-$
decay~\lcite{Aaij:2016zhm}. The original definitions of those so-called
\emph{special} angles are easily seen to be equivalent to:
\begin{equation*}
	\cos\Phi_{a} = \frac{ \vec{n_a}\cdot \vec x }
		{ \sqrt{ 1 -(\vec{n_a}\cdot \vec z)^2 } }
	,\qquad
	\sin\Phi_{a} = \frac{ \vec{n_a}\cdot \vec y }
		{ \sqrt{ 1 -(\vec{n_a}\cdot \vec z)^2 } }
	,
\end{equation*}
and similarly for $a\leftrightarrow b$. Using $(\vec x,\vec y,\vec z)^T =
R(\phi,\theta,0) (\vec{x_a},\vec{y_a},\vec{z_a})^T$,
\begin{align*}
	\begin{pmatrix} \vec x\\	\vec y\\	\vec z	\end{pmatrix}
	=
	\begin{pmatrix}
		\cos\theta\cos\phi	&\;\; -\sin\phi	&\;\; \sin\theta\cos\phi	\\
		\cos\theta\sin\phi	&\;\; \cos\phi	&\;\; \sin\theta\sin\phi	\\
		-\sin\theta		&\;\; 0		&\;\; \cos\theta
	\end{pmatrix}
	\begin{pmatrix} \vec{x_a}\\	\vec{y_a}\\	\vec{z_a}	\end{pmatrix}
	,
\end{align*}
as well as $(\vec x,\vec y,\vec z)^T = R(\pi+\phi,\pi-\theta,0)
(\vec{x_b},\vec{y_b},\vec{z_b})^T$, one derives
\begin{align*}
	&\cos\Phi_a = \frac
		{\cos\theta \cos\phi \sin\phi_a + \sin\phi \cos\phi_a}
		{\sqrt{1-\sin^2\phi_a\sin^2\theta}}
	,
	&&\sin\Phi_a = \frac
		{\cos\theta \sin\phi \sin\phi_a - \cos\phi \cos\phi_a}
		{\sqrt{1-\sin^2\phi_a\sin^2\theta}},
\\
	&\cos\Phi_b = \frac
		{\cos\theta \cos\phi \sin\phi_b - \sin\phi \cos\phi_b}
		{\sqrt{1-\sin^2\phi_b\sin^2\theta}}
	,
	&&\sin\Phi_b = \frac
		{\cos\theta \sin\phi \sin\phi_a + \cos\phi \cos\phi_b}
		{\sqrt{1-\sin^2\phi_b\sin^2\theta}}.
\end{align*}
Such angular dependences do not appear in \autoref{tab:proc2}. We therefore
stress that these four asymmetries vanish identically in the $0^{1/2}\to
a^{1/2}b^{1}\to 1^{1/2}\,2^{0} \; 3^{1/2}\,4^{1/2}$ process, when $\Lambda_b$ is
produced by the strong interaction which preserves parity. Referring to
\autoref{tab:proc1}, we note the same conclusion would also hold in $0^{1/2}\to
a^{1/2}b^{1}\to 1^{1/2}\,2^{0} \; 3^{1/2}\,4^{1/2}$ processes like
$\Lambda_b\to\Lambda\, \jpsi \to p\,\pi\, \mu^+\mu^-$. Relaxing the assumptions
of our main text, \hyperref[tab:proc1_full]{Tables~\ref{tab:proc1_full}} and
\ref{tab:proc2_full} in \autoref{sec:appendix} inform us that, in both
processes, asymmetries or moments based on $\cos\Phi_a$ and $\sin\Phi_a$ are
respectively sensitive to the $\:\im{\a1}{\a2}\alpha_a P_x\:$ and
$\;\im{\a1}{\a2}\alpha_a P_y\;$ combinations of production and decay amplitudes.
In the $0^{1/2}\to a^{1/2}b^{1}\to 1^{1/2}\,2^{0} \; 3^{1/2}\,4^{1/2}$ decay,
$\cos\Phi_b$ and $\sin\Phi_b$ asymmetries respectively provide access to
$\;\left(\im{\a1}{\b1}-\im{\a2}{\b2}\right)$ $\alpha_b P_x\;$ and
$\;\left(\im{\a1}{\b1}-\im{\a2}{\b2}\right)$ $\alpha_b P_y$. They however vanish
identically in the $0^{1/2}\to a^{1/2}b^{1}\to 1^{1/2}\,2^{0} \; 3^{0}\,4^{0}$
case.

\subsection{\T-even angular asymmetries}
With a nonvanishing \T-odd polarisation component $P_z$ produced by absorptive
parts in the production amplitudes, one could also search for \CP-odd phase
differences between decay amplitudes that have identical
strong phases through \T-even angular asymmetries. In the four-body
processes featuring an intermediate resonance $a$ of spin $1/2$, this is for
instance possible with the
\begin{align*}
	\mathcal{A}^{001}_{010} =
	\int\!\!\text{d}\Omega
	\left(
		\frac{1}{\Gamma}\frac{\text{d}\Gamma}{\text{d}\Omega}-
		\frac{1}{\bar\Gamma}\frac{\text{d}\bar\Gamma}{\text{d}\Omega}
	\right)
		&\sign\{ \cos\theta_b \sin(\phi_b) \}			,\\
	\mathcal{A}^{101}_{110} =
	\int\!\!\text{d}\Omega
	\left(
		\frac{1}{\Gamma}\frac{\text{d}\Gamma}{\text{d}\Omega}-
		\frac{1}{\bar\Gamma}\frac{\text{d}\bar\Gamma}{\text{d}\Omega}
	\right)
		&\sign\{ \cos\theta\cos\theta_b \sin(\phi_a+\phi_b) \}	,\\
	\mathcal{A}^{011}_{010} =
	\int\!\!\text{d}\Omega
	\left(
		\frac{1}{\Gamma}\frac{\text{d}\Gamma}{\text{d}\Omega}-
		\frac{1}{\bar\Gamma}\frac{\text{d}\bar\Gamma}{\text{d}\Omega}
	\right)
		&\sign\{ \cos\theta_a\cos\theta_b \sin(\phi_b) \}	,\\
	\mathcal{A}^{000}_{100} =
	\int\!\!\text{d}\Omega
	\left(
		\frac{1}{\Gamma}\frac{\text{d}\Gamma}{\text{d}\Omega}-
		\frac{1}{\bar\Gamma}\frac{\text{d}\bar\Gamma}{\text{d}\Omega}
	\right)
		&\sign\{ \sin(\phi_a) \}					,\\
	\mathcal{A}^{000}_{120} =
	\int\!\!\text{d}\Omega
	\left(
		\frac{1}{\Gamma}\frac{\text{d}\Gamma}{\text{d}\Omega}-
		\frac{1}{\bar\Gamma}\frac{\text{d}\bar\Gamma}{\text{d}\Omega}
	\right)
		&\sign\{ \sin(\phi_a+2\phi_b) \}				,
\end{align*}
asymmetries (see \hyperref[tab:proc1]{Tables~\ref{tab:proc1}} and
\ref{tab:proc2}). Only the first of these is not proportional to the asymmetry
parameter $\alpha_a$, on top of $P_z$. Many more of such asymmetries can be
constructed in the case $a$ is of spin $3/2$ and we refer the reader to the
third blocks of \hyperref[tab:proc5]{Tables~\ref{tab:proc5}} and \ref{tab:proc6}.
The third blocks of \hyperref[tab:proc_3body_2]{Tables~\ref{tab:proc_3body_2}}
to \ref{tab:proc_3body_4} are relevant for three-body processes
(\autoref{tab:proc_3body_1} has no such block). It is worth stressing here that
the polarisation of the $\Lambda_b$'s observed to decay to a $J/\psi \Lambda$
final state in the LHCb detector has been constrained to be smaller than $20\%$
at the $2.7\sigma$ level \lcite{Aaij:2013oxa}.

In principle, the above asymmetries could also be nonvanishing in the presence
of \CP\ violation in the production process, combined with strong phase
differences between decay amplitudes. This is not expected to happen when the
production process is dominated by the strong interaction but could also be
checked experimentally by measuring various asymmetries. Since \CP\ violation in
production would cause $|P_z|$ to take slightly different values in the two
conjugated processes, all the above asymmetries could potentially be
nonvanishing. Moreover, the \T-odd angular asymmetries giving access to terms
proportional to $P_z$ (in the second blocks of our tables) would then be
nonvanishing even in the absence of \CP-even phase differences between decay
amplitudes. In this sense, our tables would allow to interpret the patterns
observed in the measurement of various asymmetries.

\section{Summary}
We have studied the angular distributions of some three- and four-body decays of
spin-$1/2$ states, focusing on the discrete symmetry transformation properties
of the different contributions. Some \CP-odd asymmetries discussed in the
literature have been shown to vanish identically in the decay chains considered.
Special attention has been devoted to the two types of angular asymmetries that
could serve to access small differences of \CP-odd phases between decay
amplitudes of identical \CP-even phases. The first ones are \T-odd angular
asymmetries that are not proportional to a \T-odd initial-state polarisation
component $P_z$. The second ones are \T-even angular asymmetries proportional to
$P_z$. The latter do obviously not appear in the decay of spinless particles and
are, on the other hand, the only way to access imaginary parts of decay
amplitude interferences in the three-body decays of spinning particles (with
unmeasured final-state spins). Conversely, it was stressed that some \T-odd
angular asymmetries only give access to the imaginary parts of production---and
not decay---amplitude interferences. The \T-odd angular asymmetries sensitive to
imaginary parts of production amplitude interferences could serve to verify the
assumption of \CP\ conservation in production, without relying on nonvanishing
differences of \CP-even phases between either production or decay amplitudes. So
eventually, comparing the measured patterns of asymmetries with the expectations
provided here for specific resonant intermediate states could allow to decrypt
the dynamical nature of the process scrutinized.

\section*{Acknowledgements}
I am grateful to Yuval Grossman and Maurizio Martinelli for discussions on the
topic treated here. Together with Christophe Grojean they also provided much
valued comments on the manuscript of this paper.

\appendix
\section{Appendix}\label{sec:appendix}
We present below the distributions obtained by relaxing the hypotheses made in
the main text. When parity is violated in the production of particle $0$, its
$P_x$ and $P_y$ polarisation components can be nonvanishing. In
three-body decays with a massive vector $b^1$ appearing in the final state, the
$A_{\pm}$ amplitudes for which $\lambda_b=0$ can also be nonvanishing. Moreover,
in the $b^{1} \to 3^{1/2}\,4^{1/2}$ decay, parity violation and massive $3,4$
fermions respectively produces terms proportional to:
\begin{align*}
	\alpha_b \equiv
	&\frac	{|M_b(+1/2,-1/2)|^2 - |M_b(-1/2,+1/2)|^2}
		{|M_b(+1/2,-1/2)|^2 + |M_b(-1/2,+1/2)|^2}
	,\quad\text{and}\\
	\mu_b \equiv
	&\frac	{|M_b(+1/2,+1/2)|^2 + |M_b(-1/2,-1/2)|^2}
		{|M_b(+1/2,-1/2)|^2 + |M_b(-1/2,+1/2)|^2}.
\end{align*}
\hyperref[tab:proc1_full]{Tables~\ref{tab:proc1_full}} to
\ref{tab:proc_3body_4_full} respectively extend
\hyperref[tab:proc1]{Tables~\ref{tab:proc1}} to \ref{tab:proc_3body_4} with
these additional contributions to the kinematic distributions. There,
we used the $(\tilde{P}_x,\tilde{P}_y,\tilde{P}_z)^T \equiv
R(\phi,\theta,0)^T(P_x,P_y,P_z)^T$ polarisations along the $\vec{x_a}$,
$\vec{y_a}$, and $\vec{z_a}$ directions:
\begin{equation*}
	\begin{pmatrix}
		\tilde{P}_x\\
		\tilde{P}_y\\
		\tilde{P}_z
	\end{pmatrix}
	=
	\begin{pmatrix}
	\cos\theta\cos\phi	&\;\; \cos\theta\sin\phi	&\;\; -\sin\theta	\\
	-\sin\phi		&\;\; \cos\phi		&\;\; 0		\\
	\sin\theta\cos\phi	&\;\; \sin\theta\sin\phi	&\;\; \cos\theta
	\end{pmatrix}
	\begin{pmatrix}
		P_x\\
		P_y\\
		P_z
	\end{pmatrix}
	.
\end{equation*}
Referring to \autoref{sec:prop} where the discrete symmetry properties of the
different quantities above were derived, one sees that $\tilde{P}_x$ is
\P-even--\T-odd while $\tilde{P}_y$ and $\tilde{P}_z$ are both \P-odd--\T-even.

\end{fmffile}

\begin{raggedright}
	\bibliographystyle{JHEP}
	\bibliography{note}
\end{raggedright}

\begin{landscape}
\begin{table}[p]\center
\ensuremath{%
	\renewcommand{\arraystretch}{1.5}
	\begin{array}{@{}*{3}{c@{\;\;}}  c@{}}
	\hline\omit\\[1mm]
	+ 3/2 		&\Big(\mod{\a1} - \mod{\a2}\Big) + \Big(\mod{\a1} + \mod{\a2}\Big) \tilde{P}_z	& 	&\Big(\sin^2\theta_b + 2 \cos^2\theta_b \mu_b\Big) \cos\theta_a \alpha_a	\\
+ 3/2 		&\Big(\mod{\a1} - \mod{\a2}\Big) \tilde{P}_z + \Big(\mod{\a1} + \mod{\a2}\Big)	& 	&\Big(\sin^2\theta_b + 2 \cos^2\theta_b \mu_b\Big)	\\
- 3/4 		&\Big(\mod{\b1} - \mod{\b2}\Big) \tilde{P}_z - \Big(\mod{\b1} + \mod{\b2}\Big)	& 	&2 \cos\theta_a \cos\theta_b \alpha_a \alpha_b + \Big(1 + 2 \sin^2\theta_b \mu_b + \cos^2\theta_b\Big)	\\
+ 3/4 		&\Big(\mod{\b1} - \mod{\b2}\Big) - \Big(\mod{\b1} + \mod{\b2}\Big) \tilde{P}_z	& 	&2 \cos\theta_b \alpha_b + \Big(1 + 2 \sin^2\theta_b \mu_b + \cos^2\theta_b\Big) \cos\theta_a \alpha_a	\\
+ 3 		&\re{\a1}{\a2} \tilde{P}_x - \im{\a1}{\a2} \tilde{P}_y	&\cos\phi_a	&\Big(\sin^2\theta_b + 2 \cos^2\theta_b \mu_b\Big) \sin\theta_a \alpha_a	\\
+ 3\big/2\sqrt2 	&\Big(\re{\a1}{\b1} - \re{\a2}{\b2}\Big) \tilde{P}_x - \Big(\im{\a1}{\b1} + \im{\a2}{\b2}\Big) \tilde{P}_y	&\cos\phi_b	&2 \cos\theta_a \sin\theta_b \alpha_a \alpha_b + \Big(1 - 2 \mu_b\Big) \sin2\theta_b	\\
+ 3\big/2\sqrt2 	&\Big(\re{\a1}{\b1} + \re{\a2}{\b2}\Big) \tilde{P}_x - \Big(\im{\a1}{\b1} - \im{\a2}{\b2}\Big) \tilde{P}_y	&\cos\phi_b	&2 \sin\theta_b \alpha_b + \Big(1 - 2 \mu_b\Big) \cos\theta_a \sin2\theta_b \alpha_a	\\
+ 3/2 		&\re{\b1}{\b2} \tilde{P}_x + \im{\b1}{\b2} \tilde{P}_y	&\cos\Big(\phi_a + 2 \phi_b\Big)	&\Big(1 - 2 \mu_b\Big) \sin\theta_a \sin^2\theta_b \alpha_a	\\
- 3\big/2\sqrt2 	&\Big(\re{\a1}{\b2} - \re{\a2}{\b1}\Big) + \Big(\re{\a1}{\b2} + \re{\a2}{\b1}\Big) \tilde{P}_z	&\cos\Big(\phi_a + \phi_b\Big)	&\Big(1 - 2 \mu_b\Big) \sin\theta_a \sin2\theta_b \alpha_a	\\
+ 3 	/\sqrt2 	&\Big(\re{\a1}{\b2} - \re{\a2}{\b1}\Big) \tilde{P}_z + \Big(\re{\a1}{\b2} + \re{\a2}{\b1}\Big)	&\cos\Big(\phi_a + \phi_b\Big)	&\sin\theta_a \sin\theta_b \alpha_a \alpha_b	\\
+ 3 		&\re{\a1}{\a2} \tilde{P}_y + \im{\a1}{\a2} \tilde{P}_x	&\sin\phi_a	&\Big(\sin^2\theta_b + 2 \cos^2\theta_b \mu_b\Big) \sin\theta_a \alpha_a	\\
- 3\big/2\sqrt2 	&\Big(\re{\a1}{\b1} - \re{\a2}{\b2}\Big) \tilde{P}_y + \Big(\im{\a1}{\b1} + \im{\a2}{\b2}\Big) \tilde{P}_x	&\sin\phi_b	&2 \cos\theta_a \sin\theta_b \alpha_a \alpha_b + \Big(1 - 2 \mu_b\Big) \sin2\theta_b	\\
- 3\big/2\sqrt2 	&\Big(\re{\a1}{\b1} + \re{\a2}{\b2}\Big) \tilde{P}_y + \Big(\im{\a1}{\b1} - \im{\a2}{\b2}\Big) \tilde{P}_x	&\sin\phi_b	&2 \sin\theta_b \alpha_b + \Big(1 - 2 \mu_b\Big) \cos\theta_a \sin2\theta_b \alpha_a	\\
- 3/2 		&\re{\b1}{\b2} \tilde{P}_y - \im{\b1}{\b2} \tilde{P}_x	&\sin\Big(\phi_a + 2 \phi_b\Big)	&\Big(1 - 2 \mu_b\Big) \sin\theta_a \sin^2\theta_b \alpha_a	\\
- 3\big/2\sqrt2 	&\Big(\im{\a1}{\b2} - \im{\a2}{\b1}\Big) \tilde{P}_z + \Big(\im{\a1}{\b2} + \im{\a2}{\b1}\Big)	&\sin\Big(\phi_a + \phi_b\Big)	&\Big(1 - 2 \mu_b\Big) \sin\theta_a \sin2\theta_b \alpha_a	\\
+ 3 	/\sqrt2 	&\Big(\im{\a1}{\b2} - \im{\a2}{\b1}\Big) + \Big(\im{\a1}{\b2} + \im{\a2}{\b1}\Big) \tilde{P}_z	&\sin\Big(\phi_a + \phi_b\Big)	&\sin\theta_a \sin\theta_b \alpha_a \alpha_b	\\

	\omit\\[2mm]\hline
	\end{array}
}\caption{All contributions to $0^{1/2}\to a^{1/2}b^{1} \to 1^{1/2}\,2^{0}
          \; 3^{1/2}\,4^{1/2}$ angular distribution which appear when the
          assumptions leading to \autoref{tab:proc1} are relaxed, so that
          $P_x$, $P_y$, $\alpha_b$ or $\mu_b$ defined in the text are
          nonvanishing. We have defined $(\tilde{P}_x,\tilde{P}_y,\tilde{P}_z)
          \equiv R(\phi,\theta,0)^T(P_x,P_y,P_z)$.}
\label{tab:proc1_full}
\end{table}
\end{landscape}

\begin{table}[p]\center
\hspace*{-2.25cm}
\scalebox{.65}{
\ensuremath{%
	\renewcommand{\arraystretch}{1.5}
	\begin{array}{@{} c@{} c@{\hspace{-5mm}} c@{\quad} c@{}}
	\hline\omit\\[1mm]
	- 3/4 		&\Big(\mod{\a1} - \mod{\a2}\Big) + \Big(\mod{\a1} + \mod{\a2}\Big) \tilde{P}_z	& 	&\Big(\sin^2\theta_b + 2 \cos^2\theta_b \mu_b\Big) \Big(5 - 9 \cos^2\theta_a\Big) \cos\theta_a \alpha_a	\\
+ 3/4 		&\Big(\mod{\a1} - \mod{\a2}\Big) \tilde{P}_z + \Big(\mod{\a1} + \mod{\a2}\Big)	& 	&\Big(\sin^2\theta_b + 2 \cos^2\theta_b \mu_b\Big) \Big(1 + 3 \cos^2\theta_a\Big)	\\
+ 9/8 		&\Big(3 \mod{\b1} - 3 \mod{\b2} - \mod{\c1} + \mod{\c2}\Big) - \Big(3 \mod{\b1} + 3 \mod{\b2} + \mod{\c1} + \mod{\c2}\Big) \tilde{P}_z	& 	&\Big(1 + 2 \sin^2\theta_b \mu_b + \cos^2\theta_b\Big) \cos^3\theta_a \alpha_a	\\
- 9/8 		&\Big(\mod{\b1} - \mod{\b2} + \mod{\c1} - \mod{\c2}\Big) \tilde{P}_z - \Big(\mod{\b1} + \mod{\b2} - \mod{\c1} - \mod{\c2}\Big)	& 	&\Big(1 + 2 \sin^2\theta_b \mu_b + \cos^2\theta_b\Big) \cos^2\theta_a	\\
- 3/8 		&\Big(5 \mod{\b1} - 5 \mod{\b2} - 3 \mod{\c1} + 3 \mod{\c2}\Big) - \Big(5 \mod{\b1} + 5 \mod{\b2} + 3 \mod{\c1} + 3 \mod{\c2}\Big) \tilde{P}_z	& 	&\Big(1 + 2 \sin^2\theta_b \mu_b + \cos^2\theta_b\Big) \cos\theta_a \alpha_a	\\
- 3/8 		&\Big(\mod{\b1} - \mod{\b2} - 3 \mod{\c1} + 3 \mod{\c2}\Big) \tilde{P}_z - \Big(\mod{\b1} + \mod{\b2} + 3 \mod{\c1} + 3 \mod{\c2}\Big)	& 	&\Big(1 + 2 \sin^2\theta_b \mu_b + \cos^2\theta_b\Big)	\\
- 9/4 		&\Big(3 \mod{\b1} - 3 \mod{\b2} + \mod{\c1} - \mod{\c2}\Big) \tilde{P}_z - \Big(3 \mod{\b1} + 3 \mod{\b2} - \mod{\c1} - \mod{\c2}\Big)	& 	&\cos^3\theta_a \cos\theta_b \alpha_a \alpha_b	\\
+ 9/4 		&\Big(\mod{\b1} - \mod{\b2} - \mod{\c1} + \mod{\c2}\Big) - \Big(\mod{\b1} + \mod{\b2} + \mod{\c1} + \mod{\c2}\Big) \tilde{P}_z	& 	&\cos^2\theta_a \cos\theta_b \alpha_b	\\
+ 3/4 		&\Big(5 \mod{\b1} - 5 \mod{\b2} + 3 \mod{\c1} - 3 \mod{\c2}\Big) \tilde{P}_z - \Big(5 \mod{\b1} + 5 \mod{\b2} - 3 \mod{\c1} - 3 \mod{\c2}\Big)	& 	&\cos\theta_a \cos\theta_b \alpha_a \alpha_b	\\
+ 3/4 		&\Big(\mod{\b1} - \mod{\b2} + 3 \mod{\c1} - 3 \mod{\c2}\Big) - \Big(\mod{\b1} + \mod{\b2} - 3 \mod{\c1} - 3 \mod{\c2}\Big) \tilde{P}_z	& 	&\cos\theta_b \alpha_b	\\
- 3/2 		&\re{\a1}{\a2} \tilde{P}_x - \im{\a1}{\a2} \tilde{P}_y	&\cos\phi_a	&\Big(\sin^2\theta_b + 2 \cos^2\theta_b \mu_b\Big) \Big(1 - 9 \cos^2\theta_a\Big) \sin\theta_a \alpha_a	\\
- 3\sqrt3\big/4 	&\Big(\re{\b1}{\c1} - \re{\b2}{\c2}\Big) \tilde{P}_x + \Big(\im{\b1}{\c1} + \im{\b2}{\c2}\Big) \tilde{P}_y	&\cos\phi_a	&\Big(1 + 2 \sin^2\theta_b \mu_b + \cos^2\theta_b\Big) \sin2\theta_a - 2 \Big(1 - 3 \cos^2\theta_a\Big) \sin\theta_a \cos\theta_b \alpha_a \alpha_b	\\
- 3\sqrt3\big/4 	&\Big(\re{\b1}{\c1} + \re{\b2}{\c2}\Big) \tilde{P}_x + \Big(\im{\b1}{\c1} - \im{\b2}{\c2}\Big) \tilde{P}_y	&\cos\phi_a	&2 \sin2\theta_a \cos\theta_b \alpha_b - \Big(1 + 2 \sin^2\theta_b \mu_b + \cos^2\theta_b\Big) \Big(1 - 3 \cos^2\theta_a\Big) \sin\theta_a \alpha_a	\\
- 3\big/4\sqrt2 	&\Big(\re{\a1}{\b1} + \re{\a2}{\b2}\Big) \tilde{P}_x - \Big(\im{\a1}{\b1} - \im{\a2}{\b2}\Big) \tilde{P}_y	&\cos\phi_b	&\Big(1 - 2 \mu_b\Big) \Big(5 - 9 \cos^2\theta_a\Big) \cos\theta_a \sin2\theta_b \alpha_a - 2 \Big(1 + 3 \cos^2\theta_a\Big) \sin\theta_b \alpha_b	\\
+ 3\big/4\sqrt2 	&\Big(\re{\a1}{\b1} - \re{\a2}{\b2}\Big) \tilde{P}_x - \Big(\im{\a1}{\b1} + \im{\a2}{\b2}\Big) \tilde{P}_y	&\cos\phi_b	&\Big(1 - 2 \mu_b\Big) \Big(1 + 3 \cos^2\theta_a\Big) \sin2\theta_b - 2 \Big(5 - 9 \cos^2\theta_a\Big) \cos\theta_a \sin\theta_b \alpha_a \alpha_b	\\
- 9/4 		&\re{\c1}{\c2} \tilde{P}_x - \im{\c1}{\c2} \tilde{P}_y	&\cos\Big(3 \phi_a + 2 \phi_b\Big)	&\Big(1 - 2 \mu_b\Big) \sin\theta_a^3 \sin^2\theta_b \alpha_a	\\
+ 9\sqrt3\big/4 	&\Big(\re{\b1}{\c2} - \re{\b2}{\c1}\Big) - \Big(\re{\b1}{\c2} + \re{\b2}{\c1}\Big) \tilde{P}_z	&\cos\Big(2 \phi_a + 2 \phi_b\Big)	&\Big(1 - 2 \mu_b\Big) \sin\theta_a^2 \cos\theta_a \sin^2\theta_b \alpha_a	\\
+ 3\sqrt3\big/4 	&\Big(\re{\b1}{\c2} - \re{\b2}{\c1}\Big) \tilde{P}_z - \Big(\re{\b1}{\c2} + \re{\b2}{\c1}\Big)	&\cos\Big(2 \phi_a + 2 \phi_b\Big)	&\Big(1 - 2 \mu_b\Big) \sin\theta_a^2 \sin^2\theta_b	\\
+ 3\sqrt{2/3}\big/4 	&\Big(\re{\a1}{\c2} - \re{\a2}{\c1}\Big) \tilde{P}_x - \Big(\im{\a1}{\c2} + \im{\a2}{\c1}\Big) \tilde{P}_y	&\cos\Big(2 \phi_a + \phi_b\Big)	&6 \sin\theta_a^2 \cos\theta_a \sin\theta_b \alpha_a \alpha_b + \Big(1 - 2 \mu_b\Big) \sin\theta_a^2 \sin2\theta_b	\\
- 3\sqrt{2/3}\big/4 	&\Big(\re{\a1}{\c2} + \re{\a2}{\c1}\Big) \tilde{P}_x - \Big(\im{\a1}{\c2} - \im{\a2}{\c1}\Big) \tilde{P}_y	&\cos\Big(2 \phi_a + \phi_b\Big)	&2 \sin\theta_a^2 \sin\theta_b \alpha_b + 3 \Big(1 - 2 \mu_b\Big) \sin\theta_a^2 \cos\theta_a \sin2\theta_b \alpha_a	\\
- 3/4 		&\re{\b1}{\b2} \tilde{P}_x + \im{\b1}{\b2} \tilde{P}_y	&\cos\Big(\phi_a + 2 \phi_b\Big)	&\Big(1 - 2 \mu_b\Big) \Big(1 - 9 \cos^2\theta_a\Big) \sin\theta_a \sin^2\theta_b \alpha_a	\\
- 3\big/2\sqrt2 	&\Big(\re{\a1}{\b2} - \re{\a2}{\b1}\Big) \tilde{P}_z + \Big(\re{\a1}{\b2} + \re{\a2}{\b1}\Big)	&\cos\Big(\phi_a + \phi_b\Big)	&\Big(1 - 9 \cos^2\theta_a\Big) \sin\theta_a \sin\theta_b \alpha_a \alpha_b	\\
+ 3\big/4\sqrt2 	&\Big(\re{\a1}{\b2} - \re{\a2}{\b1}\Big) + \Big(\re{\a1}{\b2} + \re{\a2}{\b1}\Big) \tilde{P}_z	&\cos\Big(\phi_a + \phi_b\Big)	&\Big(1 - 2 \mu_b\Big) \Big(1 - 9 \cos^2\theta_a\Big) \sin\theta_a \sin2\theta_b \alpha_a	\\
- 3\sqrt{2/3}\big/4 	&\Big(\re{\a1}{\c1} - \re{\a2}{\c2}\Big) \tilde{P}_z + \Big(\re{\a1}{\c1} + \re{\a2}{\c2}\Big)	&\cos\Big(\phi_a + \phi_b\Big)	&\Big(1 - 2 \mu_b\Big) \sin2\theta_a \sin2\theta_b - 2 \Big(1 - 3 \cos^2\theta_a\Big) \sin\theta_a \sin\theta_b \alpha_a \alpha_b	\\
- 3\sqrt{2/3}\big/4 	&\Big(\re{\a1}{\c1} - \re{\a2}{\c2}\Big) + \Big(\re{\a1}{\c1} + \re{\a2}{\c2}\Big) \tilde{P}_z	&\cos\Big(\phi_a + \phi_b\Big)	&2 \sin2\theta_a \sin\theta_b \alpha_b - \Big(1 - 2 \mu_b\Big) \Big(1 - 3 \cos^2\theta_a\Big) \sin\theta_a \sin2\theta_b \alpha_a	\\
- 3/2 		&\re{\a1}{\a2} \tilde{P}_y + \im{\a1}{\a2} \tilde{P}_x	&\sin\phi_a	&\Big(\sin^2\theta_b + 2 \cos^2\theta_b \mu_b\Big) \Big(1 - 9 \cos^2\theta_a\Big) \sin\theta_a \alpha_a	\\
- 3\sqrt3\big/4 	&\Big(\re{\b1}{\c1} - \re{\b2}{\c2}\Big) \tilde{P}_y - \Big(\im{\b1}{\c1} + \im{\b2}{\c2}\Big) \tilde{P}_x	&\sin\phi_a	&\Big(1 + 2 \sin^2\theta_b \mu_b + \cos^2\theta_b\Big) \sin2\theta_a - 2 \Big(1 - 3 \cos^2\theta_a\Big) \sin\theta_a \cos\theta_b \alpha_a \alpha_b	\\
- 3\sqrt3\big/4 	&\Big(\re{\b1}{\c1} + \re{\b2}{\c2}\Big) \tilde{P}_y - \Big(\im{\b1}{\c1} - \im{\b2}{\c2}\Big) \tilde{P}_x	&\sin\phi_a	&2 \sin2\theta_a \cos\theta_b \alpha_b - \Big(1 + 2 \sin^2\theta_b \mu_b + \cos^2\theta_b\Big) \Big(1 - 3 \cos^2\theta_a\Big) \sin\theta_a \alpha_a	\\
+ 3\big/4\sqrt2 	&\Big(\re{\a1}{\b1} + \re{\a2}{\b2}\Big) \tilde{P}_y + \Big(\im{\a1}{\b1} - \im{\a2}{\b2}\Big) \tilde{P}_x	&\sin\phi_b	&\Big(1 - 2 \mu_b\Big) \Big(5 - 9 \cos^2\theta_a\Big) \cos\theta_a \sin2\theta_b \alpha_a - 2 \Big(1 + 3 \cos^2\theta_a\Big) \sin\theta_b \alpha_b	\\
- 3\big/4\sqrt2 	&\Big(\re{\a1}{\b1} - \re{\a2}{\b2}\Big) \tilde{P}_y + \Big(\im{\a1}{\b1} + \im{\a2}{\b2}\Big) \tilde{P}_x	&\sin\phi_b	&\Big(1 - 2 \mu_b\Big) \Big(1 + 3 \cos^2\theta_a\Big) \sin2\theta_b - 2 \Big(5 - 9 \cos^2\theta_a\Big) \cos\theta_a \sin\theta_b \alpha_a \alpha_b	\\
- 9/4 		&\re{\c1}{\c2} \tilde{P}_y + \im{\c1}{\c2} \tilde{P}_x	&\sin\Big(3 \phi_a + 2 \phi_b\Big)	&\Big(1 - 2 \mu_b\Big) \sin\theta_a^3 \sin^2\theta_b \alpha_a	\\
- 9\sqrt3\big/4 	&\Big(\im{\b1}{\c2} - \im{\b2}{\c1}\Big) \tilde{P}_z - \Big(\im{\b1}{\c2} + \im{\b2}{\c1}\Big)	&\sin\Big(2 \phi_a + 2 \phi_b\Big)	&\Big(1 - 2 \mu_b\Big) \sin\theta_a^2 \cos\theta_a \sin^2\theta_b \alpha_a	\\
- 3\sqrt3\big/4 	&\Big(\im{\b1}{\c2} - \im{\b2}{\c1}\Big) - \Big(\im{\b1}{\c2} + \im{\b2}{\c1}\Big) \tilde{P}_z	&\sin\Big(2 \phi_a + 2 \phi_b\Big)	&\Big(1 - 2 \mu_b\Big) \sin\theta_a^2 \sin^2\theta_b	\\
+ 3\sqrt{2/3}\big/4 	&\Big(\re{\a1}{\c2} - \re{\a2}{\c1}\Big) \tilde{P}_y + \Big(\im{\a1}{\c2} + \im{\a2}{\c1}\Big) \tilde{P}_x	&\sin\Big(2 \phi_a + \phi_b\Big)	&6 \sin\theta_a^2 \cos\theta_a \sin\theta_b \alpha_a \alpha_b + \Big(1 - 2 \mu_b\Big) \sin\theta_a^2 \sin2\theta_b	\\
- 3\sqrt{2/3}\big/4 	&\Big(\re{\a1}{\c2} + \re{\a2}{\c1}\Big) \tilde{P}_y + \Big(\im{\a1}{\c2} - \im{\a2}{\c1}\Big) \tilde{P}_x	&\sin\Big(2 \phi_a + \phi_b\Big)	&2 \sin\theta_a^2 \sin\theta_b \alpha_b + 3 \Big(1 - 2 \mu_b\Big) \sin\theta_a^2 \cos\theta_a \sin2\theta_b \alpha_a	\\
+ 3/4 		&\re{\b1}{\b2} \tilde{P}_y - \im{\b1}{\b2} \tilde{P}_x	&\sin\Big(\phi_a + 2 \phi_b\Big)	&\Big(1 - 2 \mu_b\Big) \Big(1 - 9 \cos^2\theta_a\Big) \sin\theta_a \sin^2\theta_b \alpha_a	\\
- 3\big/2\sqrt2 	&\Big(\im{\a1}{\b2} - \im{\a2}{\b1}\Big) + \Big(\im{\a1}{\b2} + \im{\a2}{\b1}\Big) \tilde{P}_z	&\sin\Big(\phi_a + \phi_b\Big)	&\Big(1 - 9 \cos^2\theta_a\Big) \sin\theta_a \sin\theta_b \alpha_a \alpha_b	\\
+ 3\big/4\sqrt2 	&\Big(\im{\a1}{\b2} - \im{\a2}{\b1}\Big) \tilde{P}_z + \Big(\im{\a1}{\b2} + \im{\a2}{\b1}\Big)	&\sin\Big(\phi_a + \phi_b\Big)	&\Big(1 - 2 \mu_b\Big) \Big(1 - 9 \cos^2\theta_a\Big) \sin\theta_a \sin2\theta_b \alpha_a	\\
+ 3\sqrt{2/3}\big/4 	&\Big(\im{\a1}{\c1} - \im{\a2}{\c2}\Big) + \Big(\im{\a1}{\c1} + \im{\a2}{\c2}\Big) \tilde{P}_z	&\sin\Big(\phi_a + \phi_b\Big)	&\Big(1 - 2 \mu_b\Big) \sin2\theta_a \sin2\theta_b - 2 \Big(1 - 3 \cos^2\theta_a\Big) \sin\theta_a \sin\theta_b \alpha_a \alpha_b	\\
+ 3\sqrt{2/3}\big/4 	&\Big(\im{\a1}{\c1} - \im{\a2}{\c2}\Big) \tilde{P}_z + \Big(\im{\a1}{\c1} + \im{\a2}{\c2}\Big)	&\sin\Big(\phi_a + \phi_b\Big)	&2 \sin2\theta_a \sin\theta_b \alpha_b - \Big(1 - 2 \mu_b\Big) \Big(1 - 3 \cos^2\theta_a\Big) \sin\theta_a \sin2\theta_b \alpha_a	\\

	\omit\\[2mm]\hline
	\end{array}
}
}
\caption{Same as \autoref{tab:proc1_full}, for the $0^{1/2}\to a^{3/2}b^{1} \to
         1^{1/2}\,2^{0} \; 3^{1/2}\,4^{1/2}$ process in which particle $a$ has
         spin $3/2$ instead of $1/2$. This table generalises
         \autoref{tab:proc5}.}
\label{tab:proc5_full}
\end{table}

\begin{landscape}
\begin{table}[p]\center
\ensuremath{%
	\renewcommand{\arraystretch}{1.5}
	\begin{array}{@{}*{3}{c@{\;\;}}  c@{}}
	\hline\omit\\[1mm]
	+ 3 		&\Big(\mod{\a1} - \mod{\a2}\Big) + \Big(\mod{\a1} + \mod{\a2}\Big) \tilde{P}_z	& 	&\cos\theta_a \cos^2\theta_b \alpha_a	\\
+ 3/2 		&\Big(\mod{\b1} - \mod{\b2}\Big) - \Big(\mod{\b1} + \mod{\b2}\Big) \tilde{P}_z	& 	&\cos\theta_a \sin^2\theta_b \alpha_a	\\
+ 3 		&\Big(\mod{\a1} - \mod{\a2}\Big) \tilde{P}_z + \Big(\mod{\a1} + \mod{\a2}\Big)	& 	&\cos^2\theta_b	\\
- 3/2 		&\Big(\mod{\b1} - \mod{\b2}\Big) \tilde{P}_z - \Big(\mod{\b1} + \mod{\b2}\Big)	& 	&\sin^2\theta_b	\\
+ 6 		&\re{\a1}{\a2} \tilde{P}_x - \im{\a1}{\a2} \tilde{P}_y	&\cos\phi_a	&\sin\theta_a \cos^2\theta_b \alpha_a	\\
- 3 	/\sqrt2 	&\Big(\re{\a1}{\b1} - \re{\a2}{\b2}\Big) \tilde{P}_x - \Big(\im{\a1}{\b1} + \im{\a2}{\b2}\Big) \tilde{P}_y	&\cos\phi_b	&\sin2\theta_b	\\
- 3 	/\sqrt2 	&\Big(\re{\a1}{\b1} + \re{\a2}{\b2}\Big) \tilde{P}_x - \Big(\im{\a1}{\b1} - \im{\a2}{\b2}\Big) \tilde{P}_y	&\cos\phi_b	&\cos\theta_a \sin2\theta_b \alpha_a	\\
- 3 		&\re{\b1}{\b2} \tilde{P}_x + \im{\b1}{\b2} \tilde{P}_y	&\cos\Big(\phi_a + 2 \phi_b\Big)	&\sin\theta_a \sin^2\theta_b \alpha_a	\\
+ 3 	/\sqrt2 	&\Big(\re{\a1}{\b2} - \re{\a2}{\b1}\Big) + \Big(\re{\a1}{\b2} + \re{\a2}{\b1}\Big) \tilde{P}_z	&\cos\Big(\phi_a + \phi_b\Big)	&\sin\theta_a \sin2\theta_b \alpha_a	\\
+ 6 		&\re{\a1}{\a2} \tilde{P}_y + \im{\a1}{\a2} \tilde{P}_x	&\sin\phi_a	&\sin\theta_a \cos^2\theta_b \alpha_a	\\
+ 3 	/\sqrt2 	&\Big(\re{\a1}{\b1} - \re{\a2}{\b2}\Big) \tilde{P}_y + \Big(\im{\a1}{\b1} + \im{\a2}{\b2}\Big) \tilde{P}_x	&\sin\phi_b	&\sin2\theta_b	\\
+ 3 	/\sqrt2 	&\Big(\re{\a1}{\b1} + \re{\a2}{\b2}\Big) \tilde{P}_y + \Big(\im{\a1}{\b1} - \im{\a2}{\b2}\Big) \tilde{P}_x	&\sin\phi_b	&\cos\theta_a \sin2\theta_b \alpha_a	\\
+ 3 		&\re{\b1}{\b2} \tilde{P}_y - \im{\b1}{\b2} \tilde{P}_x	&\sin\Big(\phi_a + 2 \phi_b\Big)	&\sin\theta_a \sin^2\theta_b \alpha_a	\\
+ 3 	/\sqrt2 	&\Big(\im{\a1}{\b2} - \im{\a2}{\b1}\Big) \tilde{P}_z + \Big(\im{\a1}{\b2} + \im{\a2}{\b1}\Big)	&\sin\Big(\phi_a + \phi_b\Big)	&\sin\theta_a \sin2\theta_b \alpha_a	\\

	\omit\\[2mm]\hline
	\end{array}
}
\caption{All contributions to $0^{1/2}\to a^{1/2}b^{1} \to 1^{1/2}\,2^{0}
         \; 3^{0}\,4^{0}$ angular distribution which appear with the assumptions
         leading to \autoref{tab:proc2} are relaxed, so that $P_x$ and
         $P_y$ are nonvanishing. We have defined
         $(\tilde{P}_x,\tilde{P}_y,\tilde{P}_z) \equiv
         R(\phi,\theta,0)^T(P_x,P_y,P_z)$.}
\label{tab:proc2_full}
\end{table}
\end{landscape}

\begin{table}[p]\center
\hspace*{-1.75cm}
\scalebox{.8}{%
\ensuremath{%
	\renewcommand{\arraystretch}{1.5}
	\begin{array}{@{} c@{} c@{\hspace{-10mm}} c@{\quad} c@{}}
	\hline\omit\\[1mm]
	- 3/2 		&\Big(\mod{\a1} - \mod{\a2}\Big) + \Big(\mod{\a1} + \mod{\a2}\Big) \tilde{P}_z	& 	&\Big(5 - 9 \cos^2\theta_a\Big) \cos\theta_a \cos^2\theta_b \alpha_a	\\
+ 3/2 		&\Big(\mod{\a1} - \mod{\a2}\Big) \tilde{P}_z + \Big(\mod{\a1} + \mod{\a2}\Big)	& 	&\Big(1 + 3 \cos^2\theta_a\Big) \cos^2\theta_b	\\
+ 9/4 		&\Big(3 \mod{\b1} - 3 \mod{\b2} - \mod{\c1} + \mod{\c2}\Big) - \Big(3 \mod{\b1} + 3 \mod{\b2} + \mod{\c1} + \mod{\c2}\Big) \tilde{P}_z	& 	&\cos^3\theta_a \sin^2\theta_b \alpha_a	\\
- 9/4 		&\Big(\mod{\b1} - \mod{\b2} + \mod{\c1} - \mod{\c2}\Big) \tilde{P}_z - \Big(\mod{\b1} + \mod{\b2} - \mod{\c1} - \mod{\c2}\Big)	& 	&\cos^2\theta_a \sin^2\theta_b	\\
- 3/4 		&\Big(5 \mod{\b1} - 5 \mod{\b2} - 3 \mod{\c1} + 3 \mod{\c2}\Big) - \Big(5 \mod{\b1} + 5 \mod{\b2} + 3 \mod{\c1} + 3 \mod{\c2}\Big) \tilde{P}_z	& 	&\cos\theta_a \sin^2\theta_b \alpha_a	\\
- 3/4 		&\Big(\mod{\b1} - \mod{\b2} - 3 \mod{\c1} + 3 \mod{\c2}\Big) \tilde{P}_z - \Big(\mod{\b1} + \mod{\b2} + 3 \mod{\c1} + 3 \mod{\c2}\Big)	& 	&\sin^2\theta_b	\\
+ 3\sqrt3\big/2 	&\Big(\re{\b1}{\c1} + \re{\b2}{\c2}\Big) \tilde{P}_x + \Big(\im{\b1}{\c1} - \im{\b2}{\c2}\Big) \tilde{P}_y	&\cos\phi_a	&\Big(1 - 3 \cos^2\theta_a\Big) \sin\theta_a \sin^2\theta_b \alpha_a	\\
- 3 		&\re{\a1}{\a2} \tilde{P}_x - \im{\a1}{\a2} \tilde{P}_y	&\cos\phi_a	&\Big(1 - 9 \cos^2\theta_a\Big) \sin\theta_a \cos^2\theta_b \alpha_a	\\
- 3\sqrt3\big/2 	&\Big(\re{\b1}{\c1} - \re{\b2}{\c2}\Big) \tilde{P}_x + \Big(\im{\b1}{\c1} + \im{\b2}{\c2}\Big) \tilde{P}_y	&\cos\phi_a	&\sin2\theta_a \sin^2\theta_b	\\
+ 3\big/2\sqrt2 	&\Big(\re{\a1}{\b1} + \re{\a2}{\b2}\Big) \tilde{P}_x - \Big(\im{\a1}{\b1} - \im{\a2}{\b2}\Big) \tilde{P}_y	&\cos\phi_b	&\Big(5 - 9 \cos^2\theta_a\Big) \cos\theta_a \sin2\theta_b \alpha_a	\\
- 3\big/2\sqrt2 	&\Big(\re{\a1}{\b1} - \re{\a2}{\b2}\Big) \tilde{P}_x - \Big(\im{\a1}{\b1} + \im{\a2}{\b2}\Big) \tilde{P}_y	&\cos\phi_b	&\Big(1 + 3 \cos^2\theta_a\Big) \sin2\theta_b	\\
+ 9/2 		&\re{\c1}{\c2} \tilde{P}_x - \im{\c1}{\c2} \tilde{P}_y	&\cos\Big(3 \phi_a + 2 \phi_b\Big)	&\sin\theta_a^3 \sin^2\theta_b \alpha_a	\\
- 9\sqrt3\big/2 	&\Big(\re{\b1}{\c2} - \re{\b2}{\c1}\Big) - \Big(\re{\b1}{\c2} + \re{\b2}{\c1}\Big) \tilde{P}_z	&\cos\Big(2 \phi_a + 2 \phi_b\Big)	&\sin\theta_a^2 \cos\theta_a \sin^2\theta_b \alpha_a	\\
- 3\sqrt3\big/2 	&\Big(\re{\b1}{\c2} - \re{\b2}{\c1}\Big) \tilde{P}_z - \Big(\re{\b1}{\c2} + \re{\b2}{\c1}\Big)	&\cos\Big(2 \phi_a + 2 \phi_b\Big)	&\sin\theta_a^2 \sin^2\theta_b	\\
+ 9\sqrt{2/3}\big/2 	&\Big(\re{\a1}{\c2} + \re{\a2}{\c1}\Big) \tilde{P}_x - \Big(\im{\a1}{\c2} - \im{\a2}{\c1}\Big) \tilde{P}_y	&\cos\Big(2 \phi_a + \phi_b\Big)	&\sin\theta_a^2 \cos\theta_a \sin2\theta_b \alpha_a	\\
- 3\sqrt{2/3}\big/2 	&\Big(\re{\a1}{\c2} - \re{\a2}{\c1}\Big) \tilde{P}_x - \Big(\im{\a1}{\c2} + \im{\a2}{\c1}\Big) \tilde{P}_y	&\cos\Big(2 \phi_a + \phi_b\Big)	&\sin\theta_a^2 \sin2\theta_b	\\
+ 3/2 		&\re{\b1}{\b2} \tilde{P}_x + \im{\b1}{\b2} \tilde{P}_y	&\cos\Big(\phi_a + 2 \phi_b\Big)	&\Big(1 - 9 \cos^2\theta_a\Big) \sin\theta_a \sin^2\theta_b \alpha_a	\\
- 3\sqrt{2/3}\big/2 	&\Big(\re{\a1}{\c1} - \re{\a2}{\c2}\Big) + \Big(\re{\a1}{\c1} + \re{\a2}{\c2}\Big) \tilde{P}_z	&\cos\Big(\phi_a + \phi_b\Big)	&\Big(1 - 3 \cos^2\theta_a\Big) \sin\theta_a \sin2\theta_b \alpha_a	\\
- 3\big/2\sqrt2 	&\Big(\re{\a1}{\b2} - \re{\a2}{\b1}\Big) + \Big(\re{\a1}{\b2} + \re{\a2}{\b1}\Big) \tilde{P}_z	&\cos\Big(\phi_a + \phi_b\Big)	&\Big(1 - 9 \cos^2\theta_a\Big) \sin\theta_a \sin2\theta_b \alpha_a	\\
+ 3\sqrt{2/3}\big/2 	&\Big(\re{\a1}{\c1} - \re{\a2}{\c2}\Big) \tilde{P}_z + \Big(\re{\a1}{\c1} + \re{\a2}{\c2}\Big)	&\cos\Big(\phi_a + \phi_b\Big)	&\sin2\theta_a \sin2\theta_b	\\
+ 3\sqrt3\big/2 	&\Big(\re{\b1}{\c1} + \re{\b2}{\c2}\Big) \tilde{P}_y - \Big(\im{\b1}{\c1} - \im{\b2}{\c2}\Big) \tilde{P}_x	&\sin\phi_a	&\Big(1 - 3 \cos^2\theta_a\Big) \sin\theta_a \sin^2\theta_b \alpha_a	\\
- 3 		&\re{\a1}{\a2} \tilde{P}_y + \im{\a1}{\a2} \tilde{P}_x	&\sin\phi_a	&\Big(1 - 9 \cos^2\theta_a\Big) \sin\theta_a \cos^2\theta_b \alpha_a	\\
- 3\sqrt3\big/2 	&\Big(\re{\b1}{\c1} - \re{\b2}{\c2}\Big) \tilde{P}_y - \Big(\im{\b1}{\c1} + \im{\b2}{\c2}\Big) \tilde{P}_x	&\sin\phi_a	&\sin2\theta_a \sin^2\theta_b	\\
- 3\big/2\sqrt2 	&\Big(\re{\a1}{\b1} + \re{\a2}{\b2}\Big) \tilde{P}_y + \Big(\im{\a1}{\b1} - \im{\a2}{\b2}\Big) \tilde{P}_x	&\sin\phi_b	&\Big(5 - 9 \cos^2\theta_a\Big) \cos\theta_a \sin2\theta_b \alpha_a	\\
+ 3\big/2\sqrt2 	&\Big(\re{\a1}{\b1} - \re{\a2}{\b2}\Big) \tilde{P}_y + \Big(\im{\a1}{\b1} + \im{\a2}{\b2}\Big) \tilde{P}_x	&\sin\phi_b	&\Big(1 + 3 \cos^2\theta_a\Big) \sin2\theta_b	\\
+ 9/2 		&\re{\c1}{\c2} \tilde{P}_y + \im{\c1}{\c2} \tilde{P}_x	&\sin\Big(3 \phi_a + 2 \phi_b\Big)	&\sin\theta_a^3 \sin^2\theta_b \alpha_a	\\
+ 9\sqrt3\big/2 	&\Big(\im{\b1}{\c2} - \im{\b2}{\c1}\Big) \tilde{P}_z - \Big(\im{\b1}{\c2} + \im{\b2}{\c1}\Big)	&\sin\Big(2 \phi_a + 2 \phi_b\Big)	&\sin\theta_a^2 \cos\theta_a \sin^2\theta_b \alpha_a	\\
+ 3\sqrt3\big/2 	&\Big(\im{\b1}{\c2} - \im{\b2}{\c1}\Big) - \Big(\im{\b1}{\c2} + \im{\b2}{\c1}\Big) \tilde{P}_z	&\sin\Big(2 \phi_a + 2 \phi_b\Big)	&\sin\theta_a^2 \sin^2\theta_b	\\
+ 9\sqrt{2/3}\big/2 	&\Big(\re{\a1}{\c2} + \re{\a2}{\c1}\Big) \tilde{P}_y + \Big(\im{\a1}{\c2} - \im{\a2}{\c1}\Big) \tilde{P}_x	&\sin\Big(2 \phi_a + \phi_b\Big)	&\sin\theta_a^2 \cos\theta_a \sin2\theta_b \alpha_a	\\
- 3\sqrt{2/3}\big/2 	&\Big(\re{\a1}{\c2} - \re{\a2}{\c1}\Big) \tilde{P}_y + \Big(\im{\a1}{\c2} + \im{\a2}{\c1}\Big) \tilde{P}_x	&\sin\Big(2 \phi_a + \phi_b\Big)	&\sin\theta_a^2 \sin2\theta_b	\\
- 3/2 		&\re{\b1}{\b2} \tilde{P}_y - \im{\b1}{\b2} \tilde{P}_x	&\sin\Big(\phi_a + 2 \phi_b\Big)	&\Big(1 - 9 \cos^2\theta_a\Big) \sin\theta_a \sin^2\theta_b \alpha_a	\\
+ 3\sqrt{2/3}\big/2 	&\Big(\im{\a1}{\c1} - \im{\a2}{\c2}\Big) \tilde{P}_z + \Big(\im{\a1}{\c1} + \im{\a2}{\c2}\Big)	&\sin\Big(\phi_a + \phi_b\Big)	&\Big(1 - 3 \cos^2\theta_a\Big) \sin\theta_a \sin2\theta_b \alpha_a	\\
- 3\big/2\sqrt2 	&\Big(\im{\a1}{\b2} - \im{\a2}{\b1}\Big) \tilde{P}_z + \Big(\im{\a1}{\b2} + \im{\a2}{\b1}\Big)	&\sin\Big(\phi_a + \phi_b\Big)	&\Big(1 - 9 \cos^2\theta_a\Big) \sin\theta_a \sin2\theta_b \alpha_a	\\
- 3\sqrt{2/3}\big/2 	&\Big(\im{\a1}{\c1} - \im{\a2}{\c2}\Big) + \Big(\im{\a1}{\c1} + \im{\a2}{\c2}\Big) \tilde{P}_z	&\sin\Big(\phi_a + \phi_b\Big)	&\sin2\theta_a \sin2\theta_b	\\

	\omit\\[2mm]\hline
	\end{array}%
}%
}%
\caption{Same as \autoref{tab:proc2_full}, for the $0^{1/2}\to a^{3/2}b^{1} \to
         1^{1/2}\,2^{0} \; 3^{0}\,4^{0}$ process, in which particle $a$ has spin
         $3/2$ instead of $1/2$. This table generalises \autoref{tab:proc6}.}
\label{tab:proc6_full}
\end{table}

\begin{table}\center
\hspace*{-.5cm}
\ensuremath{%
	\renewcommand{\arraystretch}{1.5}
	\begin{array}{@{}*{3}{c@{\quad}}  c@{}}
	\hline\omit\\[1mm]
	+ 		&\Big(\mod{\a1} - \mod{\a2} - \mod{\b1} + \mod{\b2}\Big) \tilde{P}_z + \Big(\mod{\a1} + \mod{\a2} + \mod{\b1} + \mod{\b2}\Big)	& 	& 	\\
+ 		&\Big(\mod{\a1} - \mod{\a2} + \mod{\b1} - \mod{\b2}\Big) + \Big(\mod{\a1} + \mod{\a2} - \mod{\b1} - \mod{\b2}\Big) \tilde{P}_z	& 	&\cos\theta_a \alpha_a	\\
+ 2 		&\re{\a1}{\a2} \tilde{P}_x - \im{\a1}{\a2} \tilde{P}_y	&\cos\phi_a	&\sin\theta_a \alpha_a	\\
+ 2 		&\re{\a1}{\a2} \tilde{P}_y + \im{\a1}{\a2} \tilde{P}_x	&\sin\phi_a	&\sin\theta_a \alpha_a	\\

	\omit\\[2mm]\hline
	\end{array}%
}%
\caption{
	All contributions to $0^{1/2}\to a^{1/2}b^{1} \to 1^{1/2}\,2^{0} \;
	b^{1}$ angular distribution which appear when the assumptions leading to
	\autoref{tab:proc_3body_1} are relaxed, so that $A_\pm$, $P_x$, $P_y$ defined
	in the text are nonvanishing.
}
\label{tab:proc_3body_1_full}
\end{table}

\begin{table}\center
\hspace*{-1.75cm}
\scalebox{.725}{%
\ensuremath{%
	\renewcommand{\arraystretch}{1.5}
	\begin{array}{@{} c@{\;} c@{\hspace{-15mm}} c@{\quad}  c@{}}
	\hline\omit\\[1mm]
	+ 1/2 		&\left(\mod{\a1} - \mod{\a2} - \mod{\b1} + \mod{\b2} + 3 \mod{\c1} - 3 \mod{\c2}\right) \tilde{P}_z + \left(\mod{\a1} + \mod{\a2} + \mod{\b1} + \mod{\b2} + 3 \mod{\c1} + 3 \mod{\c2}\right)	& 	& 	\\
+ 3/2 		&\left(3 \mod{\a1} - 3 \mod{\a2} + 3 \mod{\b1} - 3 \mod{\b2} - \mod{\c1} + \mod{\c2}\right) + \left(3 \mod{\a1} + 3 \mod{\a2} - 3 \mod{\b1} - 3 \mod{\b2} - \mod{\c1} - \mod{\c2}\right) \tilde{P}_z	& 	&\cos^3\theta_a \alpha_a	\\
+ 3/2 		&\left(\mod{\a1} - \mod{\a2} - \mod{\b1} + \mod{\b2} - \mod{\c1} + \mod{\c2}\right) \tilde{P}_z + \left(\mod{\a1} + \mod{\a2} + \mod{\b1} + \mod{\b2} - \mod{\c1} - \mod{\c2}\right)	& 	&\cos^2\theta_a	\\
- 1/2 		&\left(5 \mod{\a1} - 5 \mod{\a2} + 5 \mod{\b1} - 5 \mod{\b2} - 3 \mod{\c1} + 3 \mod{\c2}\right) + \left(5 \mod{\a1} + 5 \mod{\a2} - 5 \mod{\b1} - 5 \mod{\b2} - 3 \mod{\c1} - 3 \mod{\c2}\right) \tilde{P}_z	& 	&\cos\theta_a \alpha_a	\\
+ 	\sqrt3 	&\left(\re{\b1}{\c1} + \re{\b2}{\c2}\right) \tilde{P}_x + \left(\im{\b1}{\c1} - \im{\b2}{\c2}\right) \tilde{P}_y	&\cos\phi_a	&\left(1 - 3 \cos^2\theta_a\right) \sin\theta_a \alpha_a	\\
- 		&\re{\a1}{\a2} \tilde{P}_x - \im{\a1}{\a2} \tilde{P}_y	&\cos\phi_a	&\left(1 - 9 \cos^2\theta_a\right) \sin\theta_a \alpha_a	\\
- 2 	\sqrt3 	&\left(\re{\b1}{\c1} - \re{\b2}{\c2}\right) \tilde{P}_x + \left(\im{\b1}{\c1} + \im{\b2}{\c2}\right) \tilde{P}_y	&\cos\phi_a	&\cos\theta_a \sin\theta_a	\\
+ 	\sqrt3 	&\left(\re{\b1}{\c1} + \re{\b2}{\c2}\right) \tilde{P}_y - \left(\im{\b1}{\c1} - \im{\b2}{\c2}\right) \tilde{P}_x	&\sin\phi_a	&\left(1 - 3 \cos^2\theta_a\right) \sin\theta_a \alpha_a	\\
- 		&\re{\a1}{\a2} \tilde{P}_y + \im{\a1}{\a2} \tilde{P}_x	&\sin\phi_a	&\left(1 - 9 \cos^2\theta_a\right) \sin\theta_a \alpha_a	\\
- 2 	\sqrt3 	&\left(\re{\b1}{\c1} - \re{\b2}{\c2}\right) \tilde{P}_y - \left(\im{\b1}{\c1} + \im{\b2}{\c2}\right) \tilde{P}_x	&\sin\phi_a	&\cos\theta_a \sin\theta_a	\\

	\omit\\[2mm]\hline
	\end{array}%
}%
}%
\caption{
	Same as \autoref{tab:proc_3body_1_full}, for the $0^{1/2}\to a^{3/2}b^{1}
	\to 1^{1/2}\,2^{0} \; b^{1}$ process, in which particle $a$ has spin
	$3/2$ instead of $1/2$. This table generalises
	\autoref{tab:proc_3body_2}.}
\label{tab:proc_3body_2_full}
\end{table}

\begin{table}\center
\ensuremath{%
	\renewcommand{\arraystretch}{1.5}
	\begin{array}{@{}*{3}{c@{\quad}}  c@{}}
	\hline\omit\\[1mm]
	+ 		&\Big(\mod{\a1} - \mod{\a2}\Big) \tilde{P}_z + \Big(\mod{\a1} + \mod{\a2}\Big)	& 	& 	\\
+ 		&\Big(\mod{\a1} - \mod{\a2}\Big) + \Big(\mod{\a1} + \mod{\a2}\Big) \tilde{P}_z	& 	&\cos\theta_a \alpha_a	\\
+ 2 		&\re{\a1}{\a2} \tilde{P}_x - \im{\a1}{\a2} \tilde{P}_y	&\cos\phi_a	&\sin\theta_a \alpha_a	\\
+ 2 		&\re{\a1}{\a2} \tilde{P}_y + \im{\a1}{\a2} \tilde{P}_x	&\sin\phi_a	&\sin\theta_a \alpha_a	\\

	\omit\\[2mm]\hline
	\end{array}
}
\caption{
	All contributions to $0^{1/2}\to a^{1/2}b^{0} \to 1^{1/2}\,2^{0} \;
	b^{0}$ angular distribution which appear when the assumptions leading to
	\autoref{tab:proc_3body_3} are relaxed, so that $P_x$, $P_y$ defined in
	the text are nonvanishing.}
\label{tab:proc_3body_3_full}
\end{table}

\begin{table}\center
\ensuremath{%
	\renewcommand{\arraystretch}{1.5}
	\begin{array}{@{}*{3}{c@{\quad}}  c@{}}
	\hline\omit\\[1mm]
	- 1/2 		&\Big(\mod{\a1} - \mod{\a2}\Big) + \Big(\mod{\a1} + \mod{\a2}\Big) \tilde{P}_z	& 	&\Big(5 - 9 \cos^2\theta_a\Big) \cos\theta_a \alpha_a	\\
+ 1/2 		&\Big(\mod{\a1} - \mod{\a2}\Big) \tilde{P}_z + \Big(\mod{\a1} + \mod{\a2}\Big)	& 	&\Big(1 + 3 \cos^2\theta_a\Big)	\\
- 		&\re{\a1}{\a2} \tilde{P}_x - \im{\a1}{\a2} \tilde{P}_y	&\cos\phi_a	&\Big(1 - 9 \cos^2\theta_a\Big) \sin\theta_a \alpha_a	\\
- 		&\re{\a1}{\a2} \tilde{P}_y + \im{\a1}{\a2} \tilde{P}_x	&\sin\phi_a	&\Big(1 - 9 \cos^2\theta_a\Big) \sin\theta_a \alpha_a	\\

	\omit\\[2mm]\hline
	\end{array}
}
\caption{
	Same as \autoref{tab:proc_3body_3_full} for the $0^{1/2}\to a^{3/2}b^{0}
	\to 1^{1/2}\,2^{0} \; b^{0}$ process, in which particle $a$ has spin
	$3/2$ instead of $1/2$. This table generalises
	\autoref{tab:proc_3body_4}.}
\label{tab:proc_3body_4_full}
\end{table}

\end{document}